\documentclass[reprint,
superscriptaddress,
 amsmath,amssymb,
 aps,
pra,
]{revtex4-2}

\usepackage{graphicx}% Include figure files
\usepackage{dcolumn}% Align table columns on decimal point
\usepackage{bm}% bold math
\usepackage{braket}
\usepackage{hyperref}% add hypertext capabilities
\usepackage{cancel}
\usepackage[normalem]{ulem}
\hypersetup{colorlinks = true, allcolors = blue}

\usepackage[english]{babel}

\begin{document}

\title{Measuring {$\mathbb{Z}_2$} invariants in dimer models and cross-coupled ladders with a programmable photonic molecule}

\author{Sashank Kaushik Sridhar}
\thanks{These authors have contributed equally to this work.}
\affiliation{Department of Mechanical Engineering, University of Maryland, College Park, Maryland 20742, USA}

\author{Rohith Srikanth}%
\thanks{These authors have contributed equally to this work.}
\affiliation{%
 Department of Electrical Engineering, University of Maryland, College Park, Maryland 20742, USA
}

\author{Alexander R Miller}
 \affiliation{Department of Mechanical Engineering, University of Maryland, College Park, Maryland 20742, USA}

\author{Ferguson J McComb}
 \affiliation{Eleanor Roosevelt High School, Greenbelt, Maryland 20770, USA}

\author{Avik Dutt}
\email{avikdutt@umd.edu}
\affiliation{Department of Mechanical Engineering, University of Maryland, College Park, Maryland 20742, USA}

\affiliation{
 Institute for Physical Science and Technology, University of Maryland, College Park, Maryland 20742, USA
}%
\affiliation{
 National Quantum Laboratory (QLab) at Maryland, College Park, Maryland 20740, USA
}%

\date{\today}% It is always \today, today,

\begin{abstract}
% Topological effects are characterized by topological invariants, which are typically integer-valued, and lead to robust quantized transport channels in space, time and other degrees of freedom. The temporal channel in particular allows one to achieve higher-dimensional topological effects, by driving the system with multiple incommensurate frequencies. However, dissipation is generally detrimental to such topological effects. Here we introduce a photonic molecule subjected to multiple radio-frequency and optical drives as well as dissipation as a candidate system to observe quantized transport along Floquet synthetic dimensions, and provide preliminary experiments contrasting the topological and trivial phases. Topological energy pumping in the incommensurately modulated photonic molecule is enhanced by the driven-dissipative nature of our platform. Furthermore, we provide a path to realizing Weyl points and measuring the Berry curvature emanating from these reciprocal-space magnetic monopoles, illustrating the capabilities for higher-dimensional topological Hamiltonian simulation in this platform. Our approach enables direct $k$-space engineering of a wide variety of Hamiltonians using modulation bandwidths that are well below the free-spectral range of integrated photonic cavities.

Topological models are characterized by a quantized topological invariant and provide a description of novel phases of matter that can exhibit localized edge states, corner modes, and chiral transport. We experimentally realize  two 1-D lattices supporting symmetry-protected topology - the Su-Schrieffer-Heeger (SSH) and extended SSH models using the synthetic frequency dimension of coupled fiber ring resonators.  We introduce and experimentally demonstrate cascaded heterodyning as a technique for low-noise, single-shot winding number measurements through the mean chiral displacement and band structure measurements. Through our robust setup and detection techniques we can extend our capability to realizing 1-D ladder models, demonstrating a modified Creutz ladder with a staggered flux with each plaquette. This highly reconfigurable and compact fiber optics platform for Hamiltonian simulation, along with a low-noise detection scheme, provides a path forward for chip-scale realizations.

\end{abstract}

\maketitle

\textit{Introduction}: 
Analog Hamiltonian simulators are powerful platforms for emulating, and thereby verifying the dynamics of quantum and classical systems, through their superior control and measurement capabilities \cite{miyake_realizing_2013, aidelsburger_realization_2013, price_roadmap_2022, khanikaev_topologically_2015, ma_topological_2019, imhof_topolectrical-circuit_2018}.
% In addition to simulating the physics of a system that is otherwise difficult to compute, they allow for probing fundamental questions about the underlying physics.
While these simulators have been quite successful in probing 1-D and 2-D physics, many architectures including ultracold atoms \cite{aidelsburger_measuring_2015}, trapped ions \cite{blatt_quantum_2012}, and photonic resonators \cite{hafezi_imaging_2013}, run into bottlenecks set by experimental complexity and the dimensionality of the Hamiltonian, when trying to observe higher-dimensional many-body physics. One approach to bypass this is by mapping the desired spatial or temporal dynamics onto internal degrees of freedom, such as spin, orbital angular momentum, time and frequency \cite{yuan_synthetic_2021,ozawa_topological_2019,yuan_creating_2020,leefmans_topological_2022,bartlett_deterministic_2021,lustig_photonic_2019,dutt_single_2020,boada_quantum_2012,stuhl_visualizing_2015,mancini_observation_2015,sundar_synthetic_2018,luo_quantum_2015,kanungo_realizing_2022,yang_realization_2023,hu_realization_2020}. 
This new concept of `synthetic' dimensions, particularly in photonic systems, has promulgated the study of higher-order symmetries \cite{dutt_higher-order_2020}, particularly in topological lattice models, which exhibit unique properties like localized edge states
and symmetry-protected transport. Photonic systems have become the widely popular choice for topological Hamiltonian simulation owing to their versatility in encoding information,
and their relative ease of design and fabrication.

The simplest topological models, 1-D dimer models, manifest edge modes that are topologically protected via the bulk-edge correspondence \cite{asboth_short_2016}.
However, these modes are complicated to measure in infinite lattices, and lattices with periodic or even gradual boundaries, such as frequency-synthetic dimensions in ring resonators with dispersion \cite{shan_one-way_2020}. The key topological invariant in such 1-D lattices is the winding number $\mathcal{W}$, which takes on distinctly quantized values for the topological ($\mathcal{W}=1$) or trivial ($\mathcal{W}=0$) phases of the system. Many studies have thus focused on measuring the topological winding number directly in the bulk \cite{atala_direct_2013, aidelsburger_measuring_2015, zeuner_observation_2015, cardano_detection_2017}
, but these protocols often involve sophisticated dynamic measurements or slow long-time evolution.

Recently, synthetic frequency lattices have offered a solution for measuring the winding number via the mean chiral displacement, in the simplest topological lattice – the Su-Schrieffer-Heeger (SSH) model, allowing for simpler steady-state single-shot measurements. While the SSH Hamiltonian is chirally symmetric, adding an extra coupling between neighboring sites on the same sublattice breaks the chiral symmetr, but retains inversion symmetry of the SSH lattice to give the extended SSH (xSSH) lattice, which has also been theoretically shown to preserve the mean chiral displacement. The quantized mean chiral displacement finally breaks down when a synthetic magnetic flux is introduced, which can be engineered by adding a relative phase between adjacent hoppings on the lattice. 

With this flexibility to engineer phase-differences between adjacent paths, we begin to approach quasi-2D `ladder' models, whose behaviour now depends on the magnetic flux piercing each plaquette of the lattice. In 1999, Creutz \cite{creutz_end_1999}
put forth a ladder model that has been instrumental in the study of chiral transport in 2-D topological insulators, where he introduced diagonal couplings to go beyond a square lattice with magnetic flux \(\phi\) per plaquette, giving rise to strong localization due to path interference (Aharonov-Bohm caging). Experiments realizing such diagonally-coupled ladder models have been demonstrated in non-photonic systems using cold atoms \cite{kang_creutz_2020}
and a superconducting cavity \cite{hung_quantum_2021}. Nevertheless, both of these systems could benefit from simultaneously achieving fully controllable site-to-site tunneling or multiple flux plaquettes respectively, and these remain outstanding challenges.

In this Letter, we present a universal platform for realizing topological dimer models in 1D, and report experimental measurements of their topological invariants. We leverage the mode hybridisation of two identical, coupled fiber ring resonators (known as a photonic molecule) to simulate topological dimer models along a synthetic frequency dimension. Our ability to engineer a phase-difference between paths, combined with high-SNR temporal and frequency-domain measurements through cascaded optical and RF-heterodyne detection, allow us to perform single-shot band structure and site-resolved measurements of universal 1-D dimer models, thereby probing their topology. To illustrate this, we show several models of increasing complexity: the SSH model, xSSH model, and an alternating-flux analog of the Creutz ladder. Our measurements, theory and rigorous numerical simulations show excellent agreement, setting forth this platform as a valuable simulator of 1-D dimer lattices and 2-D ladder models.
\begin{figure*}[ht!]
    \centering
    \includegraphics[width = 15.5cm]{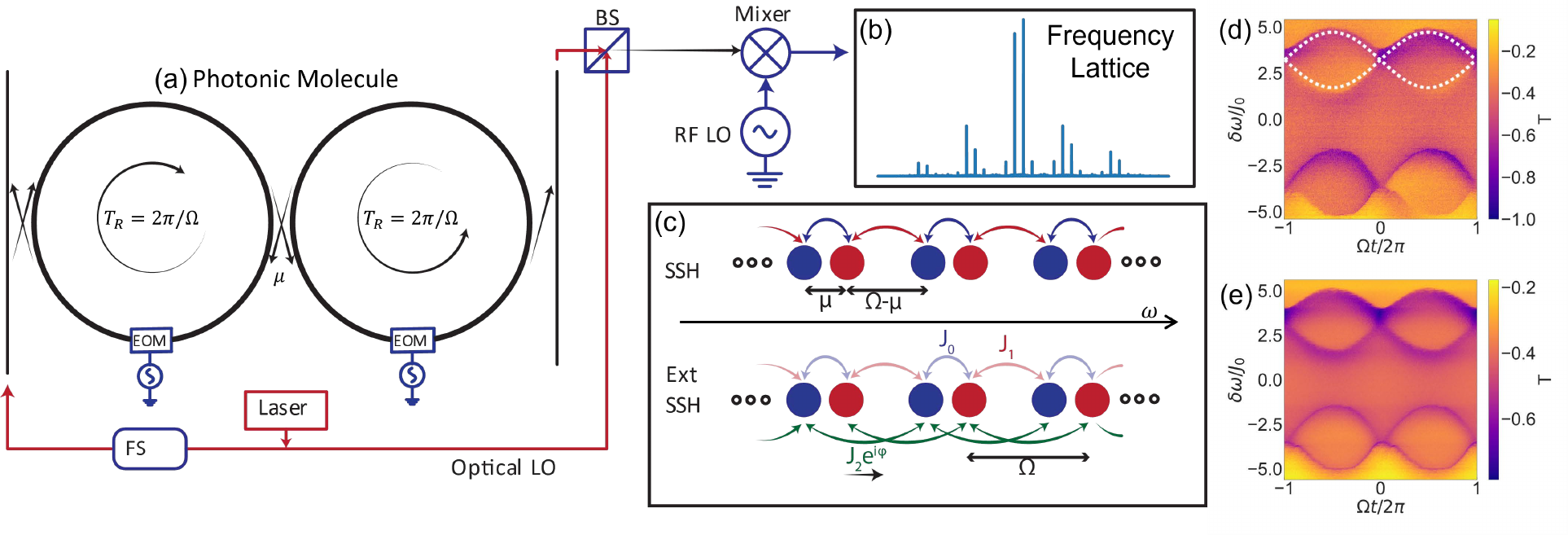}
    \caption{:(a) Schematic of a photonic molecule with electro-optic modulators (EOMs) in each ring to realize the lattices in (c). The photonic molecule supports many resonant modes, with alternating spacings $\mu$ and $\Omega-\mu$. $\Omega$ is the FSR and $\mu$ is the frequency splitting between the hybridized supermodes of the molecule. These are measured by applying a 5 GHz frequency-shift (FS) to the input laser, and performing optical and RF heterodyne at the drop port, to obtain (b), which is captured by a real-time RF spectrum analyzer (RSA). (c) Frequency lattices of the SSH and extended SSH models (xSSH). \(J_{0,1,2}\) denote EOM-induced coupling strengths at frequency separations \(\mu,\Omega - \mu\) and \(\Omega\) respectively. The xSSH model incorporates an extra coupling (green) within each sub-lattice (red/blue circles) to break chiral symmetry while preserving inversion symmetry, with an added phase $\varphi$ through the RF tone to impart a synthetic magnetic flux that also breaks inversion symmetry. (d) Band structure measurement of the SSH at the critical point \(J_1=J_0\). (e) Band structure in (d) averaged over multiple periods in k-space. Dashed white lines denote theoretical bands. BS: beam splitter. 
    }
    \label{fig:Schematic}
\end{figure*}

\textit{1-D Dimer Lattices}: The prototypical Su-Schrieffer-Heeger model \cite{su_solitons_1979} consists of a one-dimensional lattice of dimers, labelled by two distinct sub-lattices `A' and `B', with two distinct coupling strengths - intra-dimer coupling $J_0$ and inter-dimer coupling $J_1$. The topological phase transition can be observed by tuning the relative strength between the two coupling terms (topological: $J_0<J_1$, trivial: $J_0>J_1$). The extended-SSH (xSSH) model \cite{li_topological_2014} adds a next-nearest neighbor coupling $J_2$, providing a new control knob for transport across the same sub-lattice.
The tight-binding Hamiltonians for these models are:
\begin{gather}
H_{SSH}=\sum_n J_0c_{n,b}^\dagger c_{n,a}+J_1c_{n+1,a}^\dagger c_{n,b} + h.c\\
H_{\mathrm{x}SSH}=H_{SSH} + \sum_n J_2(c_{n+1,a}^\dagger c_{n,a}+c_{n+1,b}^\dagger c_{n,b})+h.c
\end{gather}
where \(c_{m,a/b}^\dagger\) (\(c_{m,a/b}\)) represents the bosonic creation (annihilation) operators, with \(m\in \mathbb{Z}\) representing the lattice site number and \(a,b\) is the sublattice index, (\(J_0,J_1, J_2\)) represent the corresponding hopping amplitudes, as is illustrated in Fig. \ref{fig:Schematic}. 

While topological protection is traditionally probed through the bulk-boundary correspondence \cite{ozawa_anomalous_2014, bardyn_measuring_2014, wimmer_experimental_2017, gianfrate_measurement_2020}, the topological winding number ($\mathcal{W}$) of the bulk lattice can be measured through a \textit{mean chiral displacement} (MCD) in chiral topological models \cite{cardano_detection_2017}. Furthermore, it has been shown theoretically by Villa \textit{et al.} \cite{Villa_meanchiral_2024} that when such models are implemented in synthetic frequency dimensions of coupled ring-resonators, the MCD can be extracted with a single-shot measurement,
\begin{equation}\label{eq:winding}
    {\langle \Gamma x\rangle }_{{{{{\rm{int}}}}}}= \frac{\mathcal{W}}{2} =\frac{\int\,d\delta \omega \,{\sum }_{n}\,n\,(| {a}_{n}^{ss}{| }^{2}-| {b}_{n}^{ss}{| }^{2})}{\int\,d\delta \omega \,{\sum }_{n}\,(| {a}_{n}^{ss}{| }^{2}+| {b}_{n}^{ss}{| }^{2})}
\end{equation}
We hence put this protocol to use for measuring the MCD as a probe of topology in frequency lattices, which do not possess well-defined boundaries and are typically measured at steady state.
\begin{figure}[ht!]
    % \centering
    \includegraphics[width=7.5cm]{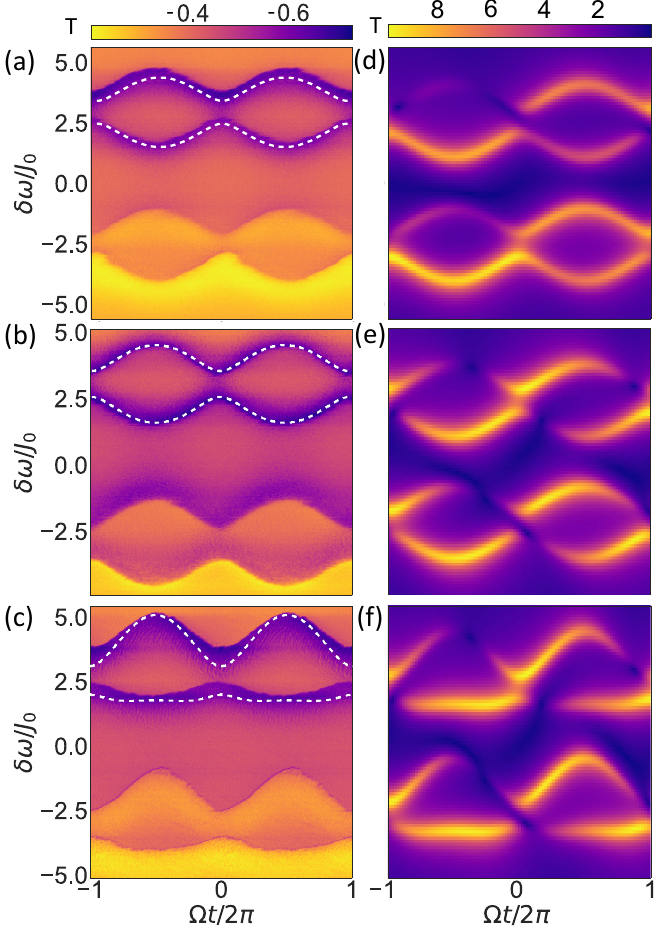}
    \caption{(a,d), (b,e), (c,f) Averaged experimental (left) and numerical (right) band structures for three  coupling regimes: (a,d) SSH lattice with \(J_1=0.5J_0\) (trivial), (b,e) SSH lattice with \(J_1=2J_0\) (topological), and (c,f) xSSH lattice with \(J_2 = 0.4J_0, J_1=2J_0\) (topological). Numerical simulations are obtained using a multi-tone dynamical coupled-mode solver, and white dotted lines represent analytical predictions. The bright and dark features of the measured bands also contain the lattice's phase information \cite{li_direct_2023, pellerin_wavefunction_2024}.
    }
    \label{fig:BS}
\end{figure} 
\begin{figure}[ht!]
    \includegraphics[scale=0.525]{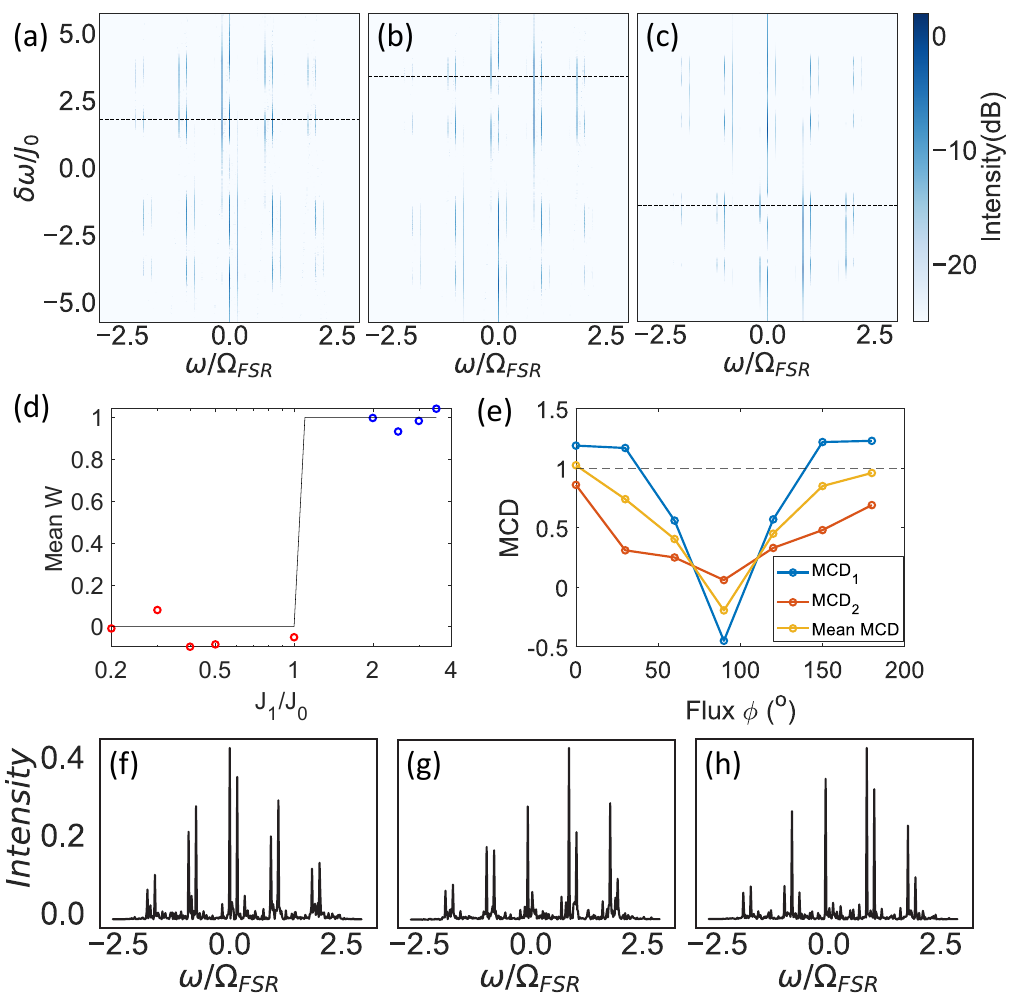}
        \caption{The optical heterodyne signal is measured on an RSA and internally mixed with an RF local oscillator (LO) to obtain in-phase (I) and quadrature (Q) time traces. The site occupations can be obtained by segmenting the full signal \(I+iQ\) into slices equal to the round trip time, stacking them vertically. The site occupations of the SSH model in the (a) trivial phase, (b) topological phase and (c) the xSSH in the topological phase are shown. (d) Winding number measurements of the SSH model for different coupling ratios (\(J_1/J_0\)). There is a qualitative distinction in the measured $\mathcal{W}$ on either side of the critical point (\(J_1=J_0\)). (e) Mean chiral displacement (MCD) measurements of the xSSH for different fluxes show a breakdown and revival of the physical correspondence between $\mathcal{W}$ and the mean chiral displacement due to loss of inversion symmetry  for $\phi \ne 0, \pi$. (f), (g) and (h) show the site occupations at specific laser detunings indicated by the linecuts in (a), (b) and (c) highlighting the emergent asymmetry in sub-lattice occupations for the topological cases and the shift in the peak of the site occupation to nonzero values.}
    \label{fig:IQ}
\end{figure}

\textit{Synthetic lattice setup}: The experiment (Fig.~\ref{fig:Schematic}), consists of two coupled fiber ring resonators with a free spectral range of \(\Omega=2\pi\times33.017\) MHz each and half-width half-maximum linewidth of $\gamma=2\pi\times364$ kHz, corresponding to a finesse $\mathcal{F}=45$ when coupled. The coupler between the rings ensures a dimer splitting of $\mu=2\pi\times6.0$ MHz averaged over 10 frequency sites. The input of the photonic molecule is 
frequency-shifted by \(5\) GHz, in order to perform coherent heterodyne detection with the same laser as a local oscillator (LO). Each ring consists of a semiconductor optical amplifier (SOA), polarization controllers and electro-optic modulators (EOMs). The SOAs and polarization controllers serve to maximize the optical power resonating at each round-trip by countering optical losses and polarization drifts. The resulting resonant modes are dimerized around each single-ring resonance, due to the creation of symmetric (+) and anti-symmetric (-) coupled-ring supermodes, allowing us to probe SSH and xSSH lattices with either supermode forming a sub-lattice across frequency space. The EOMs are driven at frequencies corresponding to the lattice separations, facilitating the required hopping terms in our Hamiltonian \cite{dutt_higher-order_2020}
. The coupling strengths $J_0,\,J_1,\,J_2$ corresponding to the applied voltages at respective frequency tones $\mu,\,\Omega-\mu$ and $\Omega$, can be independently controlled due to the unequally-spaced sites. Moreover, the voltages have been chosen such that the coupling strengths are larger than the linewidth for steady state operation, and smaller than the corresponding frequency separations they couple across ($\mu, \Omega-\mu,\Omega$), to suppress counter-rotating terms. 

\textit{Band structure measurements}: To measure the band structure, the heterodyne signal is captured using a fast photodetector and sent to an oscilloscope. The time-resolved transmission of the supermodes is captured when the EOMs are appropriately modulated. The transmission signal is broken up into slices of round trip times (fast-time), acting as the effective $k$ or crystal momentum of the lattice. The time slices are stacked on top of each other in raster-scan fashion, and the laser sweep rate (slow-time) is correlated to the laser detuning $\delta\omega$, giving us a band structure with a corresponding $E$ and $k$ variable \cite{dutt_experimental_2019-1}. With this protocol, we are able to measure the band structures of the SSH-model in its trivial and topological phases, and also provide the first direct measurement of the xSSH-model band structure on a synthetic lattice, as seen in Fig.~\ref{fig:BS}. Our experimental observations are in excellent agreement with the theoretical predictions (white dotted traces in Fig. \ref{fig:BS}(a)-(c)). We also developed a multi-tone dynamical coupled-mode simulator (Fig. \ref{fig:BS}(d)-(f)), which reproduces the bands, band-gaps, and high visibility seen in the experimentally measured bands. The brightness variations present in the simulated bands carry information about the lattice wave-function and the winding number \cite{li_direct_2023, pellerin_wavefunction_2024}, and this is also evidenced in the raw band structure data (Fig.~\ref{fig:Schematic}(d)). These features vanish when averaging over multiple fast-time periods for clearer bands, as illustrated by the data (Fig. \ref{fig:BS}(a)-(c)). Moreover, we rely on the direct observation of the frequency lattice site-occupation for ascertaining the winding number of the lattice through the MCD, the protocol for which we delineate below.

\textit{Site-resolved measurement of $\mathcal{W}$}: We obtain a high signal-to-noise ratio (SNR) site-occupation measurement through a cascaded heterodyne protocol. The fast photodiode measurement of the optical heterodyne signal is sent to a 43.5-GHz real-time RF spectrum analyser (RSA) for in-phase (I) and quadrature (Q) detection via RF heterodyne, while sweeping the laser detuning $\delta\omega$. The RSA records the I and Q time traces, thereby capturing both amplitude and phase encoded in \(I+iQ\). This data is segmented into sufficiently long time slices, stacking them vertically along the detuning axis $\delta\omega$, and then taking a Fourier transform along the synthetic lattice momentum ($\Omega t$), to obtain the \textit{frequency} lattice site occupations as a spectrogram, as seen in Fig.~\ref{fig:IQ}. The maximum instantaneous (real-time) IQ bandwidth provided by the RSA is 165 MHz, allowing us to take a single-shot measurement of 10 lattice sites (5 dimers spaced by $\Omega/2\pi = 33$ MHz). From this steady-state site occupation, we obtain the mean chiral displacement averaged over $\delta\omega$ to compute the winding numbers of the bands. For the SSH model, we measure the bulk winding number over a range of coupling ratios \(J_1/J_0\). We observe a mean value close to 0 for \(J_1/J_0\leq 1\) and a value of approximately 1 for \(J_1/J_0>1\). This remarkably clean qualitative contrast between trivial and topological phases is enabled by the high-SNR site-resolved measurements obtained through RF intermediate-frequency (IF) downconversion and subsequent IQ-detection. Although the individual windings of each set of bands is observed to be asymmetric, the mean remains consistently quantized. This hints at an imbalance between the supermodes due to the cavities not being perfectly locked to each other, in addition to slight polarization mismatch. These factors are, however, quite systematic and still preserve the overall physics in the presence of photon loss.

While this approach to computing the bulk winding number typically stands on the pillar of chiral symmetry protection, the additional coupling added to construct the xSSH lattice breaks this symmetry. However, Longhi and collaborators \cite{longhi_probing_2018, jiao_experimentally_2021} have shown that inversion symmetry is sufficient to ensure that the mean chiral displacement remains a valid observable that correlates to the winding number at steady state. This is experimentally verified by measuring the winding number for the xSSH lattice for \(J_2 = 0.4J_0, J_1 = 2J_0\) and continue to see near-perfect quantized values of $\mathcal{W}$. Furthermore, we demonstrate breakdown of this correspondence between the MCD and $\mathcal{W}$ when inversion symmetry is broken, by introducing a nonzero flux into the system (Fig.~\ref{fig:IQ}(e)). This is done by adding a phase to the $J_2$ coupling to create triangular plaquettes of equal and opposite flux in the lattice between adjacent dimers. Note that even for small values of phase, we see a drastic drop in the measured value of $\mathcal{W}$, with a subsequent revival as the phase approaches $\pi$ and restores inversion symmetry. This elucidates the utility of our technique for calibrating synthetic lattices directly through the bulk, in order to initialize topological phases at will.

\begin{figure}[ht!]   
    
    \includegraphics[width = 7.5cm]{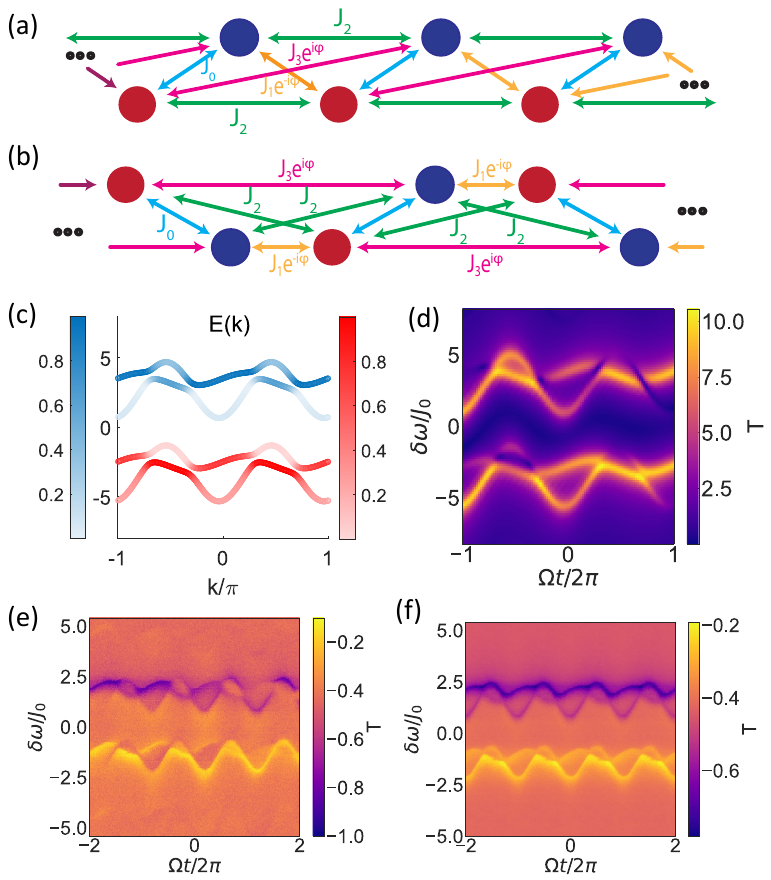}
        \caption{(a) Ladder configuration of the dimer lattice with all possible couplings within a plaquette, and individually controllable phases for the diagonal couplings. When alternate rungs of the ladder (dimers) are ``twisted", we get (b) a Creutz ladder with equal and opposite flux in adjacent plaquettes. (c) Analytical band structure for \(H(k)\), for \(J_0=J_1=J_2=J_3, \varphi=3\pi/4\). The blue (red) colour bar indicates the eigenstate projection for each value of \(k\) to the \(\sigma_x\) (\(\sigma_y\)) eigen-basis, which matches the brightness variation in (d) the simulated bands and (e) the experimental data without averaging.}
    \label{fig:creutz}
\end{figure}

While topological dimer models in frequency synthetic dimensions provide us with these robust methods to probe topological invariants, the same apparatus can then be extended to 1-D ladder models \cite{dutt_single_2020}, which exhibit the one-dimensional limits of two-dimensional topological lattices. Ladder models provide great insights into many topological phenomena such as edge transport \cite{hugel_chiral_2014, stuhl_visualizing_2015, dutt_single_2020} and AB-caging \cite{vidal_aharanov_1998}, and it thus behooves us to probe such ladder models within the same synthetic lattice framework due to the flexibility in hopping strengths and phases we can engineer in our system.

\textit{Alternating Flux Creutz Ladder}: By virtue of operating in the supermode basis of the coupled ring resonators, our system is capable of realizing a lattice model incorporating both diagonal couplings with full control over the amplitude and phase. By adding the second diagonal coupling to the previously demonstrated xSSH lattice we take a step towards realizing the Creutz ladder in synthetic frequency dimensions. However, since the couplings on both legs of the ladder are equally spaced in frequency we cannot impart a flux through those hoppings. We leverage the individual control over the diagonal couplings to address this challenge. Through the phase imparted to the corresponding drives of these couplings we can achieve a flux piercing each plaquette of the ladder. This results in a flux that alternates in sign with each plaquette \cite{anderson_staggered_2003, aidelsburger_experimental_2013, struck_engineering_2013, jotzu_experimental_2014, le_double_2024}, which can be extended to higher dimensions with higher-order RF tones \cite{senanian_programmable_2023, cheng_multidimensional_2023}.

The Hamiltonian for this lattice is described by the following equation.
\begin{gather}
H=\sum_n J_0c_{n,b}^\dagger c_{n,a} + J_2(c_{n+1,a}^\dagger c_{n,a}+c_{n+1,b}^\dagger c_{n,b}) + \\J_1e^{i\varphi}c_{n+1,b}^\dagger c_{n,a} + J_2e^{-i\varphi}c_{n+1,a}^\dagger c_{n,b} + h.c
\end{gather}

We measure the band structure of the alternating flux Creutz ladder for various coupling ratios. In Fig. \ref{fig:creutz} we show the case of equal hopping amplitudes (\(J_3 = J_2 = J_1 = J_0\)) and a flux per plaquette of \(\phi = \pm\pi/2\). Our experimental data corroborates very well with our simulations. Furthermore, we plot the analytical bands for the same parameters as shown in \ref{fig:creutz}(a). The color bars here indicate a projection of the \(k\)-space eigenstate to the \(\sigma_x\) and \(\sigma_y\) eigenbasis, which is mirrored closely by the intensity of the bands and how they evolve through the simulated and measured data. A small phase offset \(\varphi_1\) for $J_2$ is set to a non-zero value in the analytical and simulated bands to account for an uncalibrated phase on the ladder leg hopping in experiments, leading to a clear asymmetric tilt about the k=0 point. This provides the groundwork for achieving higher-dimensional topological effects and asymmetric transport in synthetic lattices of dynamically-modulated photonic ring resonators.
\textit{Conclusions}: In this work we have utilized the supermodes of a fiber-based photonic molecule to demonstrate a variety of dimer models ranging from the 1-D SSH model to a ladder based alternating flux Creutz ladder. In particular, the control and flexibility available over the RF drives allows us to shift the complexity from the device design to signal engineering, resulting in a more scalable system that can be moved to on-chip platforms. The capability to perform site-resolved measurements, along with the ability to impose boundaries \cite{dutt_creating_2022}, create unequally-spaced lattice geometries \cite{chenier_quantized_2024} and measure band structures, now places the synthetic dimensions platform at par with leading photonic Hamiltonian simulator platforms in real and synthetic space. Furthermore, the high-SNR site-resolved measurements via cascaded heterodyne and RF IQ detection provide a path forward for chip-scale implementations of our experiment, which have traditionally exhibited lower SNR \cite{dinh_reconfigurable_2024, balcytis_synthetic_2022}.
\section*{Acknowledgements}
This work was supported by an NSF CAREER award (2340835), and a Northrop Grumman seed grant. We thank Grégory Moille for support with the RF spectrum analyzer measurements.

\bibliography{My_Library, mainbib}

\providecommand{\noopsort}[1]{}\providecommand{\singleletter}[1]{#1}%
\begin{thebibliography}{58}%
\makeatletter
\providecommand \@ifxundefined [1]{%
 \@ifx{#1\undefined}
}%
\providecommand \@ifnum [1]{%
 \ifnum #1\expandafter \@firstoftwo
 \else \expandafter \@secondoftwo
 \fi
}%
\providecommand \@ifx [1]{%
 \ifx #1\expandafter \@firstoftwo
 \else \expandafter \@secondoftwo
 \fi
}%
\providecommand \natexlab [1]{#1}%
\providecommand \enquote  [1]{``#1''}%
\providecommand \bibnamefont  [1]{#1}%
\providecommand \bibfnamefont [1]{#1}%
\providecommand \citenamefont [1]{#1}%
\providecommand \href@noop [0]{\@secondoftwo}%
\providecommand \href [0]{\begingroup \@sanitize@url \@href}%
\providecommand \@href[1]{\@@startlink{#1}\@@href}%
\providecommand \@@href[1]{\endgroup#1\@@endlink}%
\providecommand \@sanitize@url [0]{\catcode `\\12\catcode `\$12\catcode `\&12\catcode `\#12\catcode `\^12\catcode `\_12\catcode `\%12\relax}%
\providecommand \@@startlink[1]{}%
\providecommand \@@endlink[0]{}%
\providecommand \url  [0]{\begingroup\@sanitize@url \@url }%
\providecommand \@url [1]{\endgroup\@href {#1}{\urlprefix }}%
\providecommand \urlprefix  [0]{URL }%
\providecommand \Eprint [0]{\href }%
\providecommand \doibase [0]{https://doi.org/}%
\providecommand \selectlanguage [0]{\@gobble}%
\providecommand \bibinfo  [0]{\@secondoftwo}%
\providecommand \bibfield  [0]{\@secondoftwo}%
\providecommand \translation [1]{[#1]}%
\providecommand \BibitemOpen [0]{}%
\providecommand \bibitemStop [0]{}%
\providecommand \bibitemNoStop [0]{.\EOS\space}%
\providecommand \EOS [0]{\spacefactor3000\relax}%
\providecommand \BibitemShut  [1]{\csname bibitem#1\endcsname}%
\let\auto@bib@innerbib\@empty
%</preamble>
\bibitem [{\citenamefont {Miyake}\ \emph {et~al.}(2013)\citenamefont {Miyake}, \citenamefont {Siviloglou}, \citenamefont {Kennedy}, \citenamefont {Burton},\ and\ \citenamefont {Ketterle}}]{miyake_realizing_2013}%
  \BibitemOpen
  \bibfield  {author} {\bibinfo {author} {\bibfnamefont {H.}~\bibnamefont {Miyake}}, \bibinfo {author} {\bibfnamefont {G.~A.}\ \bibnamefont {Siviloglou}}, \bibinfo {author} {\bibfnamefont {C.~J.}\ \bibnamefont {Kennedy}}, \bibinfo {author} {\bibfnamefont {W.~C.}\ \bibnamefont {Burton}},\ and\ \bibinfo {author} {\bibfnamefont {W.}~\bibnamefont {Ketterle}},\ }\bibfield  {title} {\bibinfo {title} {Realizing the harper hamiltonian with laser-assisted tunneling in optical lattices},\ }\href {https://doi.org/10.1103/PhysRevLett.111.185302} {\bibfield  {journal} {\bibinfo  {journal} {Phys. Rev. Lett.}\ }\textbf {\bibinfo {volume} {111}},\ \bibinfo {pages} {185302} (\bibinfo {year} {2013})}\BibitemShut {NoStop}%
\bibitem [{\citenamefont {Aidelsburger}\ \emph {et~al.}(2013{\natexlab{a}})\citenamefont {Aidelsburger}, \citenamefont {Atala}, \citenamefont {Lohse}, \citenamefont {Barreiro}, \citenamefont {Paredes},\ and\ \citenamefont {Bloch}}]{aidelsburger_realization_2013}%
  \BibitemOpen
  \bibfield  {author} {\bibinfo {author} {\bibfnamefont {M.}~\bibnamefont {Aidelsburger}}, \bibinfo {author} {\bibfnamefont {M.}~\bibnamefont {Atala}}, \bibinfo {author} {\bibfnamefont {M.}~\bibnamefont {Lohse}}, \bibinfo {author} {\bibfnamefont {J.~T.}\ \bibnamefont {Barreiro}}, \bibinfo {author} {\bibfnamefont {B.}~\bibnamefont {Paredes}},\ and\ \bibinfo {author} {\bibfnamefont {I.}~\bibnamefont {Bloch}},\ }\bibfield  {title} {\bibinfo {title} {Realization of the hofstadter hamiltonian with ultracold atoms in optical lattices},\ }\href {https://doi.org/10.1103/PhysRevLett.111.185301} {\bibfield  {journal} {\bibinfo  {journal} {Phys. Rev. Lett.}\ }\textbf {\bibinfo {volume} {111}},\ \bibinfo {pages} {185301} (\bibinfo {year} {2013}{\natexlab{a}})}\BibitemShut {NoStop}%
\bibitem [{\citenamefont {Price}\ \emph {et~al.}(2022)\citenamefont {Price}, \citenamefont {Chong}, \citenamefont {Khanikaev}, \citenamefont {Schomerus}, \citenamefont {Maczewsky}, \citenamefont {Kremer}, \citenamefont {Heinrich}, \citenamefont {Szameit}, \citenamefont {Zilberberg}, \citenamefont {Yang}, \citenamefont {Zhang}, \citenamefont {Alù}, \citenamefont {Thomale}, \citenamefont {Carusotto}, \citenamefont {St-Jean}, \citenamefont {Amo}, \citenamefont {Dutt}, \citenamefont {Yuan}, \citenamefont {Fan}, \citenamefont {Yin}, \citenamefont {Peng}, \citenamefont {Ozawa},\ and\ \citenamefont {Blanco-Redondo}}]{price_roadmap_2022}%
  \BibitemOpen
  \bibfield  {author} {\bibinfo {author} {\bibfnamefont {H.}~\bibnamefont {Price}}, \bibinfo {author} {\bibfnamefont {Y.}~\bibnamefont {Chong}}, \bibinfo {author} {\bibfnamefont {A.}~\bibnamefont {Khanikaev}}, \bibinfo {author} {\bibfnamefont {H.}~\bibnamefont {Schomerus}}, \bibinfo {author} {\bibfnamefont {L.~J.}\ \bibnamefont {Maczewsky}}, \bibinfo {author} {\bibfnamefont {M.}~\bibnamefont {Kremer}}, \bibinfo {author} {\bibfnamefont {M.}~\bibnamefont {Heinrich}}, \bibinfo {author} {\bibfnamefont {A.}~\bibnamefont {Szameit}}, \bibinfo {author} {\bibfnamefont {O.}~\bibnamefont {Zilberberg}}, \bibinfo {author} {\bibfnamefont {Y.}~\bibnamefont {Yang}}, \bibinfo {author} {\bibfnamefont {B.}~\bibnamefont {Zhang}}, \bibinfo {author} {\bibfnamefont {A.}~\bibnamefont {Alù}}, \bibinfo {author} {\bibfnamefont {R.}~\bibnamefont {Thomale}}, \bibinfo {author} {\bibfnamefont {I.}~\bibnamefont {Carusotto}}, \bibinfo {author} {\bibfnamefont {P.}~\bibnamefont {St-Jean}}, \bibinfo {author} {\bibfnamefont {A.}~\bibnamefont
  {Amo}}, \bibinfo {author} {\bibfnamefont {A.}~\bibnamefont {Dutt}}, \bibinfo {author} {\bibfnamefont {L.}~\bibnamefont {Yuan}}, \bibinfo {author} {\bibfnamefont {S.}~\bibnamefont {Fan}}, \bibinfo {author} {\bibfnamefont {X.}~\bibnamefont {Yin}}, \bibinfo {author} {\bibfnamefont {C.}~\bibnamefont {Peng}}, \bibinfo {author} {\bibfnamefont {T.}~\bibnamefont {Ozawa}},\ and\ \bibinfo {author} {\bibfnamefont {A.}~\bibnamefont {Blanco-Redondo}},\ }\bibfield  {title} {\bibinfo {title} {Roadmap on topological photonics},\ }\href {https://doi.org/10.1088/2515-7647/ac4ee4} {\bibfield  {journal} {\bibinfo  {journal} {J. Phys. Photonics}\ }\textbf {\bibinfo {volume} {4}},\ \bibinfo {pages} {032501} (\bibinfo {year} {2022})}\BibitemShut {NoStop}%
\bibitem [{\citenamefont {Khanikaev}\ \emph {et~al.}(2015)\citenamefont {Khanikaev}, \citenamefont {Fleury}, \citenamefont {Mousavi},\ and\ \citenamefont {Alu}}]{khanikaev_topologically_2015}%
  \BibitemOpen
  \bibfield  {author} {\bibinfo {author} {\bibfnamefont {A.~B.}\ \bibnamefont {Khanikaev}}, \bibinfo {author} {\bibfnamefont {R.}~\bibnamefont {Fleury}}, \bibinfo {author} {\bibfnamefont {S.~H.}\ \bibnamefont {Mousavi}},\ and\ \bibinfo {author} {\bibfnamefont {A.}~\bibnamefont {Alu}},\ }\bibfield  {title} {\bibinfo {title} {Topologically robust sound propagation in an angular-momentum-biased graphene-like resonator lattice},\ }\href@noop {} {\bibfield  {journal} {\bibinfo  {journal} {Nature communications}\ }\textbf {\bibinfo {volume} {6}},\ \bibinfo {pages} {8260} (\bibinfo {year} {2015})}\BibitemShut {NoStop}%
\bibitem [{\citenamefont {Ma}\ \emph {et~al.}(2019)\citenamefont {Ma}, \citenamefont {Xiao},\ and\ \citenamefont {Chan}}]{ma_topological_2019}%
  \BibitemOpen
  \bibfield  {author} {\bibinfo {author} {\bibfnamefont {G.}~\bibnamefont {Ma}}, \bibinfo {author} {\bibfnamefont {M.}~\bibnamefont {Xiao}},\ and\ \bibinfo {author} {\bibfnamefont {C.~T.}\ \bibnamefont {Chan}},\ }\bibfield  {title} {\bibinfo {title} {Topological phases in acoustic and mechanical systems},\ }\href@noop {} {\bibfield  {journal} {\bibinfo  {journal} {Nature Reviews Physics}\ }\textbf {\bibinfo {volume} {1}},\ \bibinfo {pages} {281} (\bibinfo {year} {2019})}\BibitemShut {NoStop}%
\bibitem [{\citenamefont {Imhof}\ \emph {et~al.}(2018)\citenamefont {Imhof}, \citenamefont {Berger}, \citenamefont {Bayer}, \citenamefont {Brehm}, \citenamefont {Molenkamp}, \citenamefont {Kiessling}, \citenamefont {Schindler}, \citenamefont {Lee}, \citenamefont {Greiter}, \citenamefont {Neupert},\ and\ \citenamefont {Thomale}}]{imhof_topolectrical-circuit_2018}%
  \BibitemOpen
  \bibfield  {author} {\bibinfo {author} {\bibfnamefont {S.}~\bibnamefont {Imhof}}, \bibinfo {author} {\bibfnamefont {C.}~\bibnamefont {Berger}}, \bibinfo {author} {\bibfnamefont {F.}~\bibnamefont {Bayer}}, \bibinfo {author} {\bibfnamefont {J.}~\bibnamefont {Brehm}}, \bibinfo {author} {\bibfnamefont {L.~W.}\ \bibnamefont {Molenkamp}}, \bibinfo {author} {\bibfnamefont {T.}~\bibnamefont {Kiessling}}, \bibinfo {author} {\bibfnamefont {F.}~\bibnamefont {Schindler}}, \bibinfo {author} {\bibfnamefont {C.~H.}\ \bibnamefont {Lee}}, \bibinfo {author} {\bibfnamefont {M.}~\bibnamefont {Greiter}}, \bibinfo {author} {\bibfnamefont {T.}~\bibnamefont {Neupert}},\ and\ \bibinfo {author} {\bibfnamefont {R.}~\bibnamefont {Thomale}},\ }\bibfield  {title} {\bibinfo {title} {Topolectrical-circuit realization of topological corner modes},\ }\href {https://doi.org/10.1038/s41567-018-0246-1} {\bibfield  {journal} {\bibinfo  {journal} {Nature Physics}\ }\textbf {\bibinfo {volume} {14}},\ \bibinfo {pages} {925} (\bibinfo {year}
  {2018})}\BibitemShut {NoStop}%
\bibitem [{\citenamefont {Aidelsburger}\ \emph {et~al.}(2015)\citenamefont {Aidelsburger}, \citenamefont {Lohse}, \citenamefont {Schweizer}, \citenamefont {Atala}, \citenamefont {Barreiro}, \citenamefont {Nascimbène}, \citenamefont {Cooper}, \citenamefont {Bloch},\ and\ \citenamefont {Goldman}}]{aidelsburger_measuring_2015}%
  \BibitemOpen
  \bibfield  {author} {\bibinfo {author} {\bibfnamefont {M.}~\bibnamefont {Aidelsburger}}, \bibinfo {author} {\bibfnamefont {M.}~\bibnamefont {Lohse}}, \bibinfo {author} {\bibfnamefont {C.}~\bibnamefont {Schweizer}}, \bibinfo {author} {\bibfnamefont {M.}~\bibnamefont {Atala}}, \bibinfo {author} {\bibfnamefont {J.~T.}\ \bibnamefont {Barreiro}}, \bibinfo {author} {\bibfnamefont {S.}~\bibnamefont {Nascimbène}}, \bibinfo {author} {\bibfnamefont {N.~R.}\ \bibnamefont {Cooper}}, \bibinfo {author} {\bibfnamefont {I.}~\bibnamefont {Bloch}},\ and\ \bibinfo {author} {\bibfnamefont {N.}~\bibnamefont {Goldman}},\ }\bibfield  {title} {\bibinfo {title} {Measuring the {Chern} number of {Hofstadter} bands with ultracold bosonic atoms},\ }\href {https://doi.org/10.1038/nphys3171} {\bibfield  {journal} {\bibinfo  {journal} {Nature Physics}\ }\textbf {\bibinfo {volume} {11}},\ \bibinfo {pages} {162} (\bibinfo {year} {2015})}\BibitemShut {NoStop}%
\bibitem [{\citenamefont {Blatt}\ and\ \citenamefont {Roos}(2012)}]{blatt_quantum_2012}%
  \BibitemOpen
  \bibfield  {author} {\bibinfo {author} {\bibfnamefont {R.}~\bibnamefont {Blatt}}\ and\ \bibinfo {author} {\bibfnamefont {C.~F.}\ \bibnamefont {Roos}},\ }\bibfield  {title} {\bibinfo {title} {Quantum simulations with trapped ions},\ }\href {https://doi.org/10.1038/nphys2252} {\bibfield  {journal} {\bibinfo  {journal} {Nature Physics}\ }\textbf {\bibinfo {volume} {8}},\ \bibinfo {pages} {277} (\bibinfo {year} {2012})}\BibitemShut {NoStop}%
\bibitem [{\citenamefont {Hafezi}\ \emph {et~al.}(2013)\citenamefont {Hafezi}, \citenamefont {Mittal}, \citenamefont {Fan}, \citenamefont {Migdall},\ and\ \citenamefont {Taylor}}]{hafezi_imaging_2013}%
  \BibitemOpen
  \bibfield  {author} {\bibinfo {author} {\bibfnamefont {M.}~\bibnamefont {Hafezi}}, \bibinfo {author} {\bibfnamefont {S.}~\bibnamefont {Mittal}}, \bibinfo {author} {\bibfnamefont {J.}~\bibnamefont {Fan}}, \bibinfo {author} {\bibfnamefont {A.}~\bibnamefont {Migdall}},\ and\ \bibinfo {author} {\bibfnamefont {J.~M.}\ \bibnamefont {Taylor}},\ }\bibfield  {title} {\bibinfo {title} {Imaging topological edge states in silicon photonics},\ }\href {https://doi.org/10.1038/nphoton.2013.274} {\bibfield  {journal} {\bibinfo  {journal} {Nature Photonics}\ }\textbf {\bibinfo {volume} {7}},\ \bibinfo {pages} {1001} (\bibinfo {year} {2013})}\BibitemShut {NoStop}%
\bibitem [{\citenamefont {Yuan}\ \emph {et~al.}(2021)\citenamefont {Yuan}, \citenamefont {Dutt},\ and\ \citenamefont {Fan}}]{yuan_synthetic_2021}%
  \BibitemOpen
  \bibfield  {author} {\bibinfo {author} {\bibfnamefont {L.}~\bibnamefont {Yuan}}, \bibinfo {author} {\bibfnamefont {A.}~\bibnamefont {Dutt}},\ and\ \bibinfo {author} {\bibfnamefont {S.}~\bibnamefont {Fan}},\ }\bibfield  {title} {\bibinfo {title} {Synthetic frequency dimensions in dynamically modulated ring resonators},\ }\href {https://doi.org/10.1063/5.0056359} {\bibfield  {journal} {\bibinfo  {journal} {APL Photonics}\ }\textbf {\bibinfo {volume} {6}},\ \bibinfo {pages} {071102} (\bibinfo {year} {2021})}\BibitemShut {NoStop}%
\bibitem [{\citenamefont {Ozawa}\ and\ \citenamefont {Price}(2019)}]{ozawa_topological_2019}%
  \BibitemOpen
  \bibfield  {author} {\bibinfo {author} {\bibfnamefont {T.}~\bibnamefont {Ozawa}}\ and\ \bibinfo {author} {\bibfnamefont {H.~M.}\ \bibnamefont {Price}},\ }\bibfield  {title} {\bibinfo {title} {Topological quantum matter in synthetic dimensions},\ }\href {https://doi.org/10.1038/s42254-019-0045-3} {\bibfield  {journal} {\bibinfo  {journal} {Nature Reviews Physics}\ }\textbf {\bibinfo {volume} {1}},\ \bibinfo {pages} {349} (\bibinfo {year} {2019})}\BibitemShut {NoStop}%
\bibitem [{\citenamefont {Yuan}\ \emph {et~al.}(2020)\citenamefont {Yuan}, \citenamefont {Dutt}, \citenamefont {Qin}, \citenamefont {Fan},\ and\ \citenamefont {Chen}}]{yuan_creating_2020}%
  \BibitemOpen
  \bibfield  {author} {\bibinfo {author} {\bibfnamefont {L.}~\bibnamefont {Yuan}}, \bibinfo {author} {\bibfnamefont {A.}~\bibnamefont {Dutt}}, \bibinfo {author} {\bibfnamefont {M.}~\bibnamefont {Qin}}, \bibinfo {author} {\bibfnamefont {S.}~\bibnamefont {Fan}},\ and\ \bibinfo {author} {\bibfnamefont {X.}~\bibnamefont {Chen}},\ }\bibfield  {title} {\bibinfo {title} {Creating locally interacting {Hamiltonians} in the synthetic frequency dimension for photons},\ }\href {https://doi.org/10.1364/PRJ.396731} {\bibfield  {journal} {\bibinfo  {journal} {Photon. Res.}\ }\textbf {\bibinfo {volume} {8}},\ \bibinfo {pages} {B8} (\bibinfo {year} {2020})}\BibitemShut {NoStop}%
\bibitem [{\citenamefont {Leefmans}\ \emph {et~al.}(2022)\citenamefont {Leefmans}, \citenamefont {Dutt}, \citenamefont {Williams}, \citenamefont {Yuan}, \citenamefont {Parto}, \citenamefont {Nori}, \citenamefont {Fan},\ and\ \citenamefont {Marandi}}]{leefmans_topological_2022}%
  \BibitemOpen
  \bibfield  {author} {\bibinfo {author} {\bibfnamefont {C.}~\bibnamefont {Leefmans}}, \bibinfo {author} {\bibfnamefont {A.}~\bibnamefont {Dutt}}, \bibinfo {author} {\bibfnamefont {J.}~\bibnamefont {Williams}}, \bibinfo {author} {\bibfnamefont {L.}~\bibnamefont {Yuan}}, \bibinfo {author} {\bibfnamefont {M.}~\bibnamefont {Parto}}, \bibinfo {author} {\bibfnamefont {F.}~\bibnamefont {Nori}}, \bibinfo {author} {\bibfnamefont {S.}~\bibnamefont {Fan}},\ and\ \bibinfo {author} {\bibfnamefont {A.}~\bibnamefont {Marandi}},\ }\bibfield  {title} {\bibinfo {title} {Topological dissipation in a time-multiplexed photonic resonator network},\ }\href {https://doi.org/10.1038/s41567-021-01492-w} {\bibfield  {journal} {\bibinfo  {journal} {Nat. Phys.}\ }\textbf {\bibinfo {volume} {18}},\ \bibinfo {pages} {442} (\bibinfo {year} {2022})}\BibitemShut {NoStop}%
\bibitem [{\citenamefont {Bartlett}\ \emph {et~al.}(2021)\citenamefont {Bartlett}, \citenamefont {Dutt},\ and\ \citenamefont {Fan}}]{bartlett_deterministic_2021}%
  \BibitemOpen
  \bibfield  {author} {\bibinfo {author} {\bibfnamefont {B.}~\bibnamefont {Bartlett}}, \bibinfo {author} {\bibfnamefont {A.}~\bibnamefont {Dutt}},\ and\ \bibinfo {author} {\bibfnamefont {S.}~\bibnamefont {Fan}},\ }\bibfield  {title} {\bibinfo {title} {Deterministic photonic quantum computation in a synthetic time dimension},\ }\href {https://doi.org/10.1364/OPTICA.424258} {\bibfield  {journal} {\bibinfo  {journal} {Optica}\ }\textbf {\bibinfo {volume} {8}},\ \bibinfo {pages} {1515} (\bibinfo {year} {2021})}\BibitemShut {NoStop}%
\bibitem [{\citenamefont {Lustig}\ \emph {et~al.}(2019)\citenamefont {Lustig}, \citenamefont {Weimann}, \citenamefont {Plotnik}, \citenamefont {Lumer}, \citenamefont {Bandres}, \citenamefont {Szameit},\ and\ \citenamefont {Segev}}]{lustig_photonic_2019}%
  \BibitemOpen
  \bibfield  {author} {\bibinfo {author} {\bibfnamefont {E.}~\bibnamefont {Lustig}}, \bibinfo {author} {\bibfnamefont {S.}~\bibnamefont {Weimann}}, \bibinfo {author} {\bibfnamefont {Y.}~\bibnamefont {Plotnik}}, \bibinfo {author} {\bibfnamefont {Y.}~\bibnamefont {Lumer}}, \bibinfo {author} {\bibfnamefont {M.~A.}\ \bibnamefont {Bandres}}, \bibinfo {author} {\bibfnamefont {A.}~\bibnamefont {Szameit}},\ and\ \bibinfo {author} {\bibfnamefont {M.}~\bibnamefont {Segev}},\ }\bibfield  {title} {\bibinfo {title} {Photonic topological insulator in synthetic dimensions},\ }\href {https://doi.org/10.1038/s41586-019-0943-7} {\bibfield  {journal} {\bibinfo  {journal} {Nature}\ }\textbf {\bibinfo {volume} {567}},\ \bibinfo {pages} {356} (\bibinfo {year} {2019})}\BibitemShut {NoStop}%
\bibitem [{\citenamefont {Dutt}\ \emph {et~al.}(2020{\natexlab{a}})\citenamefont {Dutt}, \citenamefont {Lin}, \citenamefont {Yuan}, \citenamefont {Minkov}, \citenamefont {Xiao},\ and\ \citenamefont {Fan}}]{dutt_single_2020}%
  \BibitemOpen
  \bibfield  {author} {\bibinfo {author} {\bibfnamefont {A.}~\bibnamefont {Dutt}}, \bibinfo {author} {\bibfnamefont {Q.}~\bibnamefont {Lin}}, \bibinfo {author} {\bibfnamefont {L.}~\bibnamefont {Yuan}}, \bibinfo {author} {\bibfnamefont {M.}~\bibnamefont {Minkov}}, \bibinfo {author} {\bibfnamefont {M.}~\bibnamefont {Xiao}},\ and\ \bibinfo {author} {\bibfnamefont {S.}~\bibnamefont {Fan}},\ }\bibfield  {title} {\bibinfo {title} {A single photonic cavity with two independent physical synthetic dimensions},\ }\href {https://doi.org/10.1126/science.aaz3071} {\bibfield  {journal} {\bibinfo  {journal} {Science}\ }\textbf {\bibinfo {volume} {367}},\ \bibinfo {pages} {59} (\bibinfo {year} {2020}{\natexlab{a}})}\BibitemShut {NoStop}%
\bibitem [{\citenamefont {Boada}\ \emph {et~al.}(2012)\citenamefont {Boada}, \citenamefont {Celi}, \citenamefont {Latorre},\ and\ \citenamefont {Lewenstein}}]{boada_quantum_2012}%
  \BibitemOpen
  \bibfield  {author} {\bibinfo {author} {\bibfnamefont {O.}~\bibnamefont {Boada}}, \bibinfo {author} {\bibfnamefont {A.}~\bibnamefont {Celi}}, \bibinfo {author} {\bibfnamefont {J.~I.}\ \bibnamefont {Latorre}},\ and\ \bibinfo {author} {\bibfnamefont {M.}~\bibnamefont {Lewenstein}},\ }\bibfield  {title} {\bibinfo {title} {Quantum {Simulation} of an {Extra} {Dimension}},\ }\href {https://doi.org/10.1103/PhysRevLett.108.133001} {\bibfield  {journal} {\bibinfo  {journal} {Phys. Rev. Lett.}\ }\textbf {\bibinfo {volume} {108}},\ \bibinfo {pages} {133001} (\bibinfo {year} {2012})}\BibitemShut {NoStop}%
\bibitem [{\citenamefont {Stuhl}\ \emph {et~al.}(2015)\citenamefont {Stuhl}, \citenamefont {Lu}, \citenamefont {Aycock}, \citenamefont {Genkina},\ and\ \citenamefont {Spielman}}]{stuhl_visualizing_2015}%
  \BibitemOpen
  \bibfield  {author} {\bibinfo {author} {\bibfnamefont {B.~K.}\ \bibnamefont {Stuhl}}, \bibinfo {author} {\bibfnamefont {H.-I.}\ \bibnamefont {Lu}}, \bibinfo {author} {\bibfnamefont {L.~M.}\ \bibnamefont {Aycock}}, \bibinfo {author} {\bibfnamefont {D.}~\bibnamefont {Genkina}},\ and\ \bibinfo {author} {\bibfnamefont {I.~B.}\ \bibnamefont {Spielman}},\ }\bibfield  {title} {\bibinfo {title} {Visualizing edge states with an atomic {Bose} gas in the quantum {Hall} regime},\ }\href {https://doi.org/10.1126/science.aaa8515} {\bibfield  {journal} {\bibinfo  {journal} {Science}\ }\textbf {\bibinfo {volume} {349}},\ \bibinfo {pages} {1514} (\bibinfo {year} {2015})}\BibitemShut {NoStop}%
\bibitem [{\citenamefont {Mancini}\ \emph {et~al.}(2015)\citenamefont {Mancini}, \citenamefont {Pagano}, \citenamefont {Cappellini}, \citenamefont {Livi}, \citenamefont {Rider}, \citenamefont {Catani}, \citenamefont {Sias}, \citenamefont {Zoller}, \citenamefont {Inguscio}, \citenamefont {Dalmonte},\ and\ \citenamefont {Fallani}}]{mancini_observation_2015}%
  \BibitemOpen
  \bibfield  {author} {\bibinfo {author} {\bibfnamefont {M.}~\bibnamefont {Mancini}}, \bibinfo {author} {\bibfnamefont {G.}~\bibnamefont {Pagano}}, \bibinfo {author} {\bibfnamefont {G.}~\bibnamefont {Cappellini}}, \bibinfo {author} {\bibfnamefont {L.}~\bibnamefont {Livi}}, \bibinfo {author} {\bibfnamefont {M.}~\bibnamefont {Rider}}, \bibinfo {author} {\bibfnamefont {J.}~\bibnamefont {Catani}}, \bibinfo {author} {\bibfnamefont {C.}~\bibnamefont {Sias}}, \bibinfo {author} {\bibfnamefont {P.}~\bibnamefont {Zoller}}, \bibinfo {author} {\bibfnamefont {M.}~\bibnamefont {Inguscio}}, \bibinfo {author} {\bibfnamefont {M.}~\bibnamefont {Dalmonte}},\ and\ \bibinfo {author} {\bibfnamefont {L.}~\bibnamefont {Fallani}},\ }\bibfield  {title} {\bibinfo {title} {Observation of chiral edge states with neutral fermions in synthetic {Hall} ribbons},\ }\href {https://doi.org/10.1126/science.aaa8736} {\bibfield  {journal} {\bibinfo  {journal} {Science}\ }\textbf {\bibinfo {volume} {349}},\ \bibinfo {pages} {1510} (\bibinfo {year}
  {2015})}\BibitemShut {NoStop}%
\bibitem [{\citenamefont {Sundar}\ \emph {et~al.}(2018)\citenamefont {Sundar}, \citenamefont {Gadway},\ and\ \citenamefont {Hazzard}}]{sundar_synthetic_2018}%
  \BibitemOpen
  \bibfield  {author} {\bibinfo {author} {\bibfnamefont {B.}~\bibnamefont {Sundar}}, \bibinfo {author} {\bibfnamefont {B.}~\bibnamefont {Gadway}},\ and\ \bibinfo {author} {\bibfnamefont {K.~R.~A.}\ \bibnamefont {Hazzard}},\ }\bibfield  {title} {\bibinfo {title} {Synthetic dimensions in ultracold polar molecules},\ }\href {https://doi.org/10.1038/s41598-018-21699-x} {\bibfield  {journal} {\bibinfo  {journal} {Scientific Reports}\ }\textbf {\bibinfo {volume} {8}},\ \bibinfo {pages} {3422} (\bibinfo {year} {2018})}\BibitemShut {NoStop}%
\bibitem [{\citenamefont {Luo}\ \emph {et~al.}(2015)\citenamefont {Luo}, \citenamefont {Zhou}, \citenamefont {Li}, \citenamefont {Xu}, \citenamefont {Guo},\ and\ \citenamefont {Zhou}}]{luo_quantum_2015}%
  \BibitemOpen
  \bibfield  {author} {\bibinfo {author} {\bibfnamefont {X.-W.}\ \bibnamefont {Luo}}, \bibinfo {author} {\bibfnamefont {X.}~\bibnamefont {Zhou}}, \bibinfo {author} {\bibfnamefont {C.-F.}\ \bibnamefont {Li}}, \bibinfo {author} {\bibfnamefont {J.-S.}\ \bibnamefont {Xu}}, \bibinfo {author} {\bibfnamefont {G.-C.}\ \bibnamefont {Guo}},\ and\ \bibinfo {author} {\bibfnamefont {Z.-W.}\ \bibnamefont {Zhou}},\ }\bibfield  {title} {\bibinfo {title} {Quantum simulation of {2D} topological physics in a {1D} array of optical cavities},\ }\href {https://doi.org/10.1038/ncomms8704} {\bibfield  {journal} {\bibinfo  {journal} {Nature Communications}\ }\textbf {\bibinfo {volume} {6}},\ \bibinfo {pages} {7704} (\bibinfo {year} {2015})}\BibitemShut {NoStop}%
\bibitem [{\citenamefont {Kanungo}\ \emph {et~al.}(2022)\citenamefont {Kanungo}, \citenamefont {Whalen}, \citenamefont {Lu}, \citenamefont {Yuan}, \citenamefont {Dasgupta}, \citenamefont {Dunning}, \citenamefont {Hazzard},\ and\ \citenamefont {Killian}}]{kanungo_realizing_2022}%
  \BibitemOpen
  \bibfield  {author} {\bibinfo {author} {\bibfnamefont {S.~K.}\ \bibnamefont {Kanungo}}, \bibinfo {author} {\bibfnamefont {J.~D.}\ \bibnamefont {Whalen}}, \bibinfo {author} {\bibfnamefont {Y.}~\bibnamefont {Lu}}, \bibinfo {author} {\bibfnamefont {M.}~\bibnamefont {Yuan}}, \bibinfo {author} {\bibfnamefont {S.}~\bibnamefont {Dasgupta}}, \bibinfo {author} {\bibfnamefont {F.~B.}\ \bibnamefont {Dunning}}, \bibinfo {author} {\bibfnamefont {K.~R.~A.}\ \bibnamefont {Hazzard}},\ and\ \bibinfo {author} {\bibfnamefont {T.~C.}\ \bibnamefont {Killian}},\ }\bibfield  {title} {\bibinfo {title} {Realizing topological edge states with {Rydberg}-atom synthetic dimensions},\ }\href {https://doi.org/10.1038/s41467-022-28550-y} {\bibfield  {journal} {\bibinfo  {journal} {Nat Commun}\ }\textbf {\bibinfo {volume} {13}},\ \bibinfo {pages} {972} (\bibinfo {year} {2022})}\BibitemShut {NoStop}%
\bibitem [{\citenamefont {Yang}\ \emph {et~al.}(2023)\citenamefont {Yang}, \citenamefont {Zhang}, \citenamefont {Liao}, \citenamefont {Liu}, \citenamefont {Zhou}, \citenamefont {Zhou}, \citenamefont {Xu}, \citenamefont {Han}, \citenamefont {Li},\ and\ \citenamefont {Guo}}]{yang_realization_2023}%
  \BibitemOpen
  \bibfield  {author} {\bibinfo {author} {\bibfnamefont {M.}~\bibnamefont {Yang}}, \bibinfo {author} {\bibfnamefont {H.-Q.}\ \bibnamefont {Zhang}}, \bibinfo {author} {\bibfnamefont {Y.-W.}\ \bibnamefont {Liao}}, \bibinfo {author} {\bibfnamefont {Z.-H.}\ \bibnamefont {Liu}}, \bibinfo {author} {\bibfnamefont {Z.-W.}\ \bibnamefont {Zhou}}, \bibinfo {author} {\bibfnamefont {X.-X.}\ \bibnamefont {Zhou}}, \bibinfo {author} {\bibfnamefont {J.-S.}\ \bibnamefont {Xu}}, \bibinfo {author} {\bibfnamefont {Y.-J.}\ \bibnamefont {Han}}, \bibinfo {author} {\bibfnamefont {C.-F.}\ \bibnamefont {Li}},\ and\ \bibinfo {author} {\bibfnamefont {G.-C.}\ \bibnamefont {Guo}},\ }\bibfield  {title} {\bibinfo {title} {Realization of exceptional points along a synthetic orbital angular momentum dimension},\ }\href {https://doi.org/10.1126/sciadv.abp8943} {\bibfield  {journal} {\bibinfo  {journal} {Science Advances}\ }\textbf {\bibinfo {volume} {9}},\ \bibinfo {pages} {eabp8943} (\bibinfo {year} {2023})}\BibitemShut {NoStop}%
\bibitem [{\citenamefont {Hu}\ \emph {et~al.}(2020)\citenamefont {Hu}, \citenamefont {Reimer}, \citenamefont {Shams-Ansari}, \citenamefont {Zhang},\ and\ \citenamefont {Loncar}}]{hu_realization_2020}%
  \BibitemOpen
  \bibfield  {author} {\bibinfo {author} {\bibfnamefont {Y.}~\bibnamefont {Hu}}, \bibinfo {author} {\bibfnamefont {C.}~\bibnamefont {Reimer}}, \bibinfo {author} {\bibfnamefont {A.}~\bibnamefont {Shams-Ansari}}, \bibinfo {author} {\bibfnamefont {M.}~\bibnamefont {Zhang}},\ and\ \bibinfo {author} {\bibfnamefont {M.}~\bibnamefont {Loncar}},\ }\bibfield  {title} {\bibinfo {title} {Realization of high-dimensional frequency crystals in electro-optic microcombs},\ }\href {https://doi.org/10.1364/OPTICA.395114} {\bibfield  {journal} {\bibinfo  {journal} {Optica}\ }\textbf {\bibinfo {volume} {7}},\ \bibinfo {pages} {1189} (\bibinfo {year} {2020})}\BibitemShut {NoStop}%
\bibitem [{\citenamefont {Dutt}\ \emph {et~al.}(2020{\natexlab{b}})\citenamefont {Dutt}, \citenamefont {Minkov}, \citenamefont {Williamson},\ and\ \citenamefont {Fan}}]{dutt_higher-order_2020}%
  \BibitemOpen
  \bibfield  {author} {\bibinfo {author} {\bibfnamefont {A.}~\bibnamefont {Dutt}}, \bibinfo {author} {\bibfnamefont {M.}~\bibnamefont {Minkov}}, \bibinfo {author} {\bibfnamefont {I.~A.~D.}\ \bibnamefont {Williamson}},\ and\ \bibinfo {author} {\bibfnamefont {S.}~\bibnamefont {Fan}},\ }\bibfield  {title} {\bibinfo {title} {Higher-order topological insulators in synthetic dimensions},\ }\href {https://doi.org/10.1038/s41377-020-0334-8} {\bibfield  {journal} {\bibinfo  {journal} {Light: Science \& Applications}\ }\textbf {\bibinfo {volume} {9}},\ \bibinfo {pages} {131} (\bibinfo {year} {2020}{\natexlab{b}})}\BibitemShut {NoStop}%
\bibitem [{\citenamefont {Asbóth}\ \emph {et~al.}(2016)\citenamefont {Asbóth}, \citenamefont {Oroszlány},\ and\ \citenamefont {Pályi}}]{asboth_short_2016}%
  \BibitemOpen
  \bibfield  {author} {\bibinfo {author} {\bibfnamefont {J.~K.}\ \bibnamefont {Asbóth}}, \bibinfo {author} {\bibfnamefont {L.}~\bibnamefont {Oroszlány}},\ and\ \bibinfo {author} {\bibfnamefont {A.}~\bibnamefont {Pályi}},\ }\href {http://link.springer.com/10.1007/978-3-319-25607-8} {\emph {\bibinfo {title} {A {Short} {Course} on {Topological} {Insulators}}}},\ \bibinfo {series} {Lecture {Notes} in {Physics}}, Vol.\ \bibinfo {volume} {919}\ (\bibinfo  {publisher} {Springer International Publishing},\ \bibinfo {address} {Cham},\ \bibinfo {year} {2016})\BibitemShut {NoStop}%
\bibitem [{\citenamefont {Shan}\ \emph {et~al.}(2020)\citenamefont {Shan}, \citenamefont {Yu}, \citenamefont {Li}, \citenamefont {Yuan},\ and\ \citenamefont {Chen}}]{shan_one-way_2020}%
  \BibitemOpen
  \bibfield  {author} {\bibinfo {author} {\bibfnamefont {Q.}~\bibnamefont {Shan}}, \bibinfo {author} {\bibfnamefont {D.}~\bibnamefont {Yu}}, \bibinfo {author} {\bibfnamefont {G.}~\bibnamefont {Li}}, \bibinfo {author} {\bibfnamefont {L.}~\bibnamefont {Yuan}},\ and\ \bibinfo {author} {\bibfnamefont {a.~X.}\ \bibnamefont {Chen}},\ }\bibfield  {title} {{\selectlanguage {English}\bibinfo {title} {One-{Way} {Topological} {States} {Along} {Vague} {Boundaries} in {Synthetic} {Frequency} {Dimensions} {Including} {Group} {Velocity} {Dispersion} ({Invited})}},\ }\href {https://doi.org/10.2528/PIER20083101} {\bibfield  {journal} {\bibinfo  {journal} {Progress In Electromagnetics Research}\ }\textbf {\bibinfo {volume} {169}},\ \bibinfo {pages} {33} (\bibinfo {year} {2020})}\BibitemShut {NoStop}%
\bibitem [{\citenamefont {Atala}\ \emph {et~al.}(2013)\citenamefont {Atala}, \citenamefont {Aidelsburger}, \citenamefont {Barreiro}, \citenamefont {Abanin}, \citenamefont {Kitagawa}, \citenamefont {Demler},\ and\ \citenamefont {Bloch}}]{atala_direct_2013}%
  \BibitemOpen
  \bibfield  {author} {\bibinfo {author} {\bibfnamefont {M.}~\bibnamefont {Atala}}, \bibinfo {author} {\bibfnamefont {M.}~\bibnamefont {Aidelsburger}}, \bibinfo {author} {\bibfnamefont {J.~T.}\ \bibnamefont {Barreiro}}, \bibinfo {author} {\bibfnamefont {D.}~\bibnamefont {Abanin}}, \bibinfo {author} {\bibfnamefont {T.}~\bibnamefont {Kitagawa}}, \bibinfo {author} {\bibfnamefont {E.}~\bibnamefont {Demler}},\ and\ \bibinfo {author} {\bibfnamefont {I.}~\bibnamefont {Bloch}},\ }\bibfield  {title} {\bibinfo {title} {Direct measurement of the {Zak} phase in topological {Bloch} bands},\ }\href {https://doi.org/10.1038/nphys2790} {\bibfield  {journal} {\bibinfo  {journal} {Nature Physics}\ }\textbf {\bibinfo {volume} {9}},\ \bibinfo {pages} {795} (\bibinfo {year} {2013})}\BibitemShut {NoStop}%
\bibitem [{\citenamefont {Zeuner}\ \emph {et~al.}(2015)\citenamefont {Zeuner}, \citenamefont {Rechtsman}, \citenamefont {Plotnik}, \citenamefont {Lumer}, \citenamefont {Nolte}, \citenamefont {Rudner}, \citenamefont {Segev},\ and\ \citenamefont {Szameit}}]{zeuner_observation_2015}%
  \BibitemOpen
  \bibfield  {author} {\bibinfo {author} {\bibfnamefont {J.~M.}\ \bibnamefont {Zeuner}}, \bibinfo {author} {\bibfnamefont {M.~C.}\ \bibnamefont {Rechtsman}}, \bibinfo {author} {\bibfnamefont {Y.}~\bibnamefont {Plotnik}}, \bibinfo {author} {\bibfnamefont {Y.}~\bibnamefont {Lumer}}, \bibinfo {author} {\bibfnamefont {S.}~\bibnamefont {Nolte}}, \bibinfo {author} {\bibfnamefont {M.~S.}\ \bibnamefont {Rudner}}, \bibinfo {author} {\bibfnamefont {M.}~\bibnamefont {Segev}},\ and\ \bibinfo {author} {\bibfnamefont {A.}~\bibnamefont {Szameit}},\ }\bibfield  {title} {\bibinfo {title} {Observation of a topological transition in the bulk of a non-hermitian system},\ }\href {https://doi.org/10.1103/PhysRevLett.115.040402} {\bibfield  {journal} {\bibinfo  {journal} {Phys. Rev. Lett.}\ }\textbf {\bibinfo {volume} {115}},\ \bibinfo {pages} {040402} (\bibinfo {year} {2015})}\BibitemShut {NoStop}%
\bibitem [{\citenamefont {Cardano}\ \emph {et~al.}(2017)\citenamefont {Cardano}, \citenamefont {D’Errico}, \citenamefont {Dauphin}, \citenamefont {Maffei}, \citenamefont {Piccirillo}, \citenamefont {de~Lisio}, \citenamefont {De~Filippis}, \citenamefont {Cataudella}, \citenamefont {Santamato}, \citenamefont {Marrucci}, \citenamefont {Lewenstein},\ and\ \citenamefont {Massignan}}]{cardano_detection_2017}%
  \BibitemOpen
  \bibfield  {author} {\bibinfo {author} {\bibfnamefont {F.}~\bibnamefont {Cardano}}, \bibinfo {author} {\bibfnamefont {A.}~\bibnamefont {D’Errico}}, \bibinfo {author} {\bibfnamefont {A.}~\bibnamefont {Dauphin}}, \bibinfo {author} {\bibfnamefont {M.}~\bibnamefont {Maffei}}, \bibinfo {author} {\bibfnamefont {B.}~\bibnamefont {Piccirillo}}, \bibinfo {author} {\bibfnamefont {C.}~\bibnamefont {de~Lisio}}, \bibinfo {author} {\bibfnamefont {G.}~\bibnamefont {De~Filippis}}, \bibinfo {author} {\bibfnamefont {V.}~\bibnamefont {Cataudella}}, \bibinfo {author} {\bibfnamefont {E.}~\bibnamefont {Santamato}}, \bibinfo {author} {\bibfnamefont {L.}~\bibnamefont {Marrucci}}, \bibinfo {author} {\bibfnamefont {M.}~\bibnamefont {Lewenstein}},\ and\ \bibinfo {author} {\bibfnamefont {P.}~\bibnamefont {Massignan}},\ }\bibfield  {title} {\bibinfo {title} {Detection of {Zak} phases and topological invariants in a chiral quantum walk of twisted photons},\ }\href {https://doi.org/10.1038/ncomms15516} {\bibfield  {journal} {\bibinfo
  {journal} {Nature Communications}\ }\textbf {\bibinfo {volume} {8}},\ \bibinfo {pages} {15516} (\bibinfo {year} {2017})}\BibitemShut {NoStop}%
\bibitem [{\citenamefont {Creutz}(1999)}]{creutz_end_1999}%
  \BibitemOpen
  \bibfield  {author} {\bibinfo {author} {\bibfnamefont {M.}~\bibnamefont {Creutz}},\ }\bibfield  {title} {\bibinfo {title} {End states, ladder compounds, and domain-wall fermions},\ }\href {https://doi.org/10.1103/PhysRevLett.83.2636} {\bibfield  {journal} {\bibinfo  {journal} {Phys. Rev. Lett.}\ }\textbf {\bibinfo {volume} {83}},\ \bibinfo {pages} {2636} (\bibinfo {year} {1999})}\BibitemShut {NoStop}%
\bibitem [{\citenamefont {Kang}\ \emph {et~al.}(2020)\citenamefont {Kang}, \citenamefont {Han},\ and\ \citenamefont {Shin}}]{kang_creutz_2020}%
  \BibitemOpen
  \bibfield  {author} {\bibinfo {author} {\bibfnamefont {J.~H.}\ \bibnamefont {Kang}}, \bibinfo {author} {\bibfnamefont {J.~H.}\ \bibnamefont {Han}},\ and\ \bibinfo {author} {\bibfnamefont {Y.-i.}\ \bibnamefont {Shin}},\ }\bibfield  {title} {\bibinfo {title} {Creutz ladder in a resonantly shaken 1d optical lattice},\ }\href@noop {} {\bibfield  {journal} {\bibinfo  {journal} {New Journal of Physics}\ }\textbf {\bibinfo {volume} {22}},\ \bibinfo {pages} {013023} (\bibinfo {year} {2020})}\BibitemShut {NoStop}%
\bibitem [{\citenamefont {Hung}\ \emph {et~al.}(2021)\citenamefont {Hung}, \citenamefont {Busnaina}, \citenamefont {Chang}, \citenamefont {Vadiraj}, \citenamefont {Nsanzineza}, \citenamefont {Solano}, \citenamefont {Alaeian}, \citenamefont {Rico},\ and\ \citenamefont {Wilson}}]{hung_quantum_2021}%
  \BibitemOpen
  \bibfield  {author} {\bibinfo {author} {\bibfnamefont {J.~S.~C.}\ \bibnamefont {Hung}}, \bibinfo {author} {\bibfnamefont {J.~H.}\ \bibnamefont {Busnaina}}, \bibinfo {author} {\bibfnamefont {C.~W.~S.}\ \bibnamefont {Chang}}, \bibinfo {author} {\bibfnamefont {A.~M.}\ \bibnamefont {Vadiraj}}, \bibinfo {author} {\bibfnamefont {I.}~\bibnamefont {Nsanzineza}}, \bibinfo {author} {\bibfnamefont {E.}~\bibnamefont {Solano}}, \bibinfo {author} {\bibfnamefont {H.}~\bibnamefont {Alaeian}}, \bibinfo {author} {\bibfnamefont {E.}~\bibnamefont {Rico}},\ and\ \bibinfo {author} {\bibfnamefont {C.~M.}\ \bibnamefont {Wilson}},\ }\bibfield  {title} {\bibinfo {title} {Quantum {Simulation} of the {Bosonic} {Creutz} {Ladder} with a {Parametric} {Cavity}},\ }\href {http://arxiv.org/abs/2101.03926} {\bibfield  {journal} {\bibinfo  {journal} {arXiv:2101.03926}\ } (\bibinfo {year} {2021})}\BibitemShut {NoStop}%
\bibitem [{\citenamefont {Su}\ \emph {et~al.}(1979)\citenamefont {Su}, \citenamefont {Schrieffer},\ and\ \citenamefont {Heeger}}]{su_solitons_1979}%
  \BibitemOpen
  \bibfield  {author} {\bibinfo {author} {\bibfnamefont {W.~P.}\ \bibnamefont {Su}}, \bibinfo {author} {\bibfnamefont {J.~R.}\ \bibnamefont {Schrieffer}},\ and\ \bibinfo {author} {\bibfnamefont {A.~J.}\ \bibnamefont {Heeger}},\ }\bibfield  {title} {\bibinfo {title} {Solitons in {Polyacetylene}},\ }\href {https://doi.org/10.1103/PhysRevLett.42.1698} {\bibfield  {journal} {\bibinfo  {journal} {Phys. Rev. Lett.}\ }\textbf {\bibinfo {volume} {42}},\ \bibinfo {pages} {1698} (\bibinfo {year} {1979})}\BibitemShut {NoStop}%
\bibitem [{\citenamefont {Li}\ \emph {et~al.}(2014)\citenamefont {Li}, \citenamefont {Xu},\ and\ \citenamefont {Chen}}]{li_topological_2014}%
  \BibitemOpen
  \bibfield  {author} {\bibinfo {author} {\bibfnamefont {L.}~\bibnamefont {Li}}, \bibinfo {author} {\bibfnamefont {Z.}~\bibnamefont {Xu}},\ and\ \bibinfo {author} {\bibfnamefont {S.}~\bibnamefont {Chen}},\ }\bibfield  {title} {\bibinfo {title} {Topological phases of generalized {Su}-{Schrieffer}-{Heeger} models},\ }\href {https://doi.org/10.1103/PhysRevB.89.085111} {\bibfield  {journal} {\bibinfo  {journal} {Physical Review B}\ }\textbf {\bibinfo {volume} {89}},\ \bibinfo {pages} {085111} (\bibinfo {year} {2014})}\BibitemShut {NoStop}%
\bibitem [{\citenamefont {Ozawa}\ and\ \citenamefont {Carusotto}(2014)}]{ozawa_anomalous_2014}%
  \BibitemOpen
  \bibfield  {author} {\bibinfo {author} {\bibfnamefont {T.}~\bibnamefont {Ozawa}}\ and\ \bibinfo {author} {\bibfnamefont {I.}~\bibnamefont {Carusotto}},\ }\bibfield  {title} {\bibinfo {title} {Anomalous and {Quantum} {Hall} {Effects} in {Lossy} {Photonic} {Lattices}},\ }\href {https://doi.org/10.1103/PhysRevLett.112.133902} {\bibfield  {journal} {\bibinfo  {journal} {Phys. Rev. Lett.}\ }\textbf {\bibinfo {volume} {112}},\ \bibinfo {pages} {133902} (\bibinfo {year} {2014})}\BibitemShut {NoStop}%
\bibitem [{\citenamefont {Bardyn}\ \emph {et~al.}(2014)\citenamefont {Bardyn}, \citenamefont {Huber},\ and\ \citenamefont {Zilberberg}}]{bardyn_measuring_2014}%
  \BibitemOpen
  \bibfield  {author} {\bibinfo {author} {\bibfnamefont {C.-E.}\ \bibnamefont {Bardyn}}, \bibinfo {author} {\bibfnamefont {S.~D.}\ \bibnamefont {Huber}},\ and\ \bibinfo {author} {\bibfnamefont {O.}~\bibnamefont {Zilberberg}},\ }\bibfield  {title} {\bibinfo {title} {Measuring topological invariants in small photonic lattices},\ }\href {https://doi.org/10.1088/1367-2630/16/12/123013} {\bibfield  {journal} {\bibinfo  {journal} {New Journal of Physics}\ }\textbf {\bibinfo {volume} {16}},\ \bibinfo {pages} {123013} (\bibinfo {year} {2014})}\BibitemShut {NoStop}%
\bibitem [{\citenamefont {Wimmer}\ \emph {et~al.}(2017)\citenamefont {Wimmer}, \citenamefont {Price}, \citenamefont {Carusotto},\ and\ \citenamefont {Peschel}}]{wimmer_experimental_2017}%
  \BibitemOpen
  \bibfield  {author} {\bibinfo {author} {\bibfnamefont {M.}~\bibnamefont {Wimmer}}, \bibinfo {author} {\bibfnamefont {H.~M.}\ \bibnamefont {Price}}, \bibinfo {author} {\bibfnamefont {I.}~\bibnamefont {Carusotto}},\ and\ \bibinfo {author} {\bibfnamefont {U.}~\bibnamefont {Peschel}},\ }\bibfield  {title} {\bibinfo {title} {Experimental measurement of the {Berry} curvature from anomalous transport},\ }\href {https://doi.org/10.1038/nphys4050} {\bibfield  {journal} {\bibinfo  {journal} {Nature Physics}\ }\textbf {\bibinfo {volume} {13}},\ \bibinfo {pages} {545} (\bibinfo {year} {2017})}\BibitemShut {NoStop}%
\bibitem [{\citenamefont {Gianfrate}\ \emph {et~al.}(2020)\citenamefont {Gianfrate}, \citenamefont {Bleu}, \citenamefont {Dominici}, \citenamefont {Ardizzone}, \citenamefont {De~Giorgi}, \citenamefont {Ballarini}, \citenamefont {Lerario}, \citenamefont {West}, \citenamefont {Pfeiffer}, \citenamefont {Solnyshkov}, \citenamefont {Sanvitto},\ and\ \citenamefont {Malpuech}}]{gianfrate_measurement_2020}%
  \BibitemOpen
  \bibfield  {author} {\bibinfo {author} {\bibfnamefont {A.}~\bibnamefont {Gianfrate}}, \bibinfo {author} {\bibfnamefont {O.}~\bibnamefont {Bleu}}, \bibinfo {author} {\bibfnamefont {L.}~\bibnamefont {Dominici}}, \bibinfo {author} {\bibfnamefont {V.}~\bibnamefont {Ardizzone}}, \bibinfo {author} {\bibfnamefont {M.}~\bibnamefont {De~Giorgi}}, \bibinfo {author} {\bibfnamefont {D.}~\bibnamefont {Ballarini}}, \bibinfo {author} {\bibfnamefont {G.}~\bibnamefont {Lerario}}, \bibinfo {author} {\bibfnamefont {K.~W.}\ \bibnamefont {West}}, \bibinfo {author} {\bibfnamefont {L.~N.}\ \bibnamefont {Pfeiffer}}, \bibinfo {author} {\bibfnamefont {D.~D.}\ \bibnamefont {Solnyshkov}}, \bibinfo {author} {\bibfnamefont {D.}~\bibnamefont {Sanvitto}},\ and\ \bibinfo {author} {\bibfnamefont {G.}~\bibnamefont {Malpuech}},\ }\bibfield  {title} {\bibinfo {title} {Measurement of the quantum geometric tensor and of the anomalous hall drift},\ }\href {https://doi.org/10.1038/s41586-020-1989-2} {\bibfield  {journal} {\bibinfo  {journal}
  {Nature}\ }\textbf {\bibinfo {volume} {578}},\ \bibinfo {pages} {381} (\bibinfo {year} {2020})}\BibitemShut {NoStop}%
\bibitem [{\citenamefont {Villa}\ \emph {et~al.}(2024)\citenamefont {Villa}, \citenamefont {Carusotto},\ and\ \citenamefont {Ozawa}}]{Villa_meanchiral_2024}%
  \BibitemOpen
  \bibfield  {author} {\bibinfo {author} {\bibfnamefont {G.}~\bibnamefont {Villa}}, \bibinfo {author} {\bibfnamefont {I.}~\bibnamefont {Carusotto}},\ and\ \bibinfo {author} {\bibfnamefont {T.}~\bibnamefont {Ozawa}},\ }\bibfield  {title} {\bibinfo {title} {Mean-chiral displacement in coherently driven photonic lattices and its application to synthetic frequency dimensions},\ }\href {https://doi.org/10.1038/s42005-024-01727-1} {\bibfield  {journal} {\bibinfo  {journal} {Communications Physics}\ }\textbf {\bibinfo {volume} {7}},\ \bibinfo {pages} {246} (\bibinfo {year} {2024})}\BibitemShut {NoStop}%
\bibitem [{\citenamefont {Li}\ \emph {et~al.}(2023)\citenamefont {Li}, \citenamefont {Wang}, \citenamefont {Ye}, \citenamefont {Zheng}, \citenamefont {Wang}, \citenamefont {Liu}, \citenamefont {Dutt}, \citenamefont {Yuan},\ and\ \citenamefont {Chen}}]{li_direct_2023}%
  \BibitemOpen
  \bibfield  {author} {\bibinfo {author} {\bibfnamefont {G.}~\bibnamefont {Li}}, \bibinfo {author} {\bibfnamefont {L.}~\bibnamefont {Wang}}, \bibinfo {author} {\bibfnamefont {R.}~\bibnamefont {Ye}}, \bibinfo {author} {\bibfnamefont {Y.}~\bibnamefont {Zheng}}, \bibinfo {author} {\bibfnamefont {D.-W.}\ \bibnamefont {Wang}}, \bibinfo {author} {\bibfnamefont {X.-J.}\ \bibnamefont {Liu}}, \bibinfo {author} {\bibfnamefont {A.}~\bibnamefont {Dutt}}, \bibinfo {author} {\bibfnamefont {L.}~\bibnamefont {Yuan}},\ and\ \bibinfo {author} {\bibfnamefont {X.}~\bibnamefont {Chen}},\ }\bibfield  {title} {\bibinfo {title} {Direct extraction of topological zak phase with the synthetic dimension},\ }\href {https://doi.org/10.1038/s41377-023-01126-1} {\bibfield  {journal} {\bibinfo  {journal} {Light: Science {\&} Applications}\ }\textbf {\bibinfo {volume} {12}},\ \bibinfo {pages} {81} (\bibinfo {year} {2023})}\BibitemShut {NoStop}%
\bibitem [{\citenamefont {Pellerin}\ \emph {et~al.}(2024)\citenamefont {Pellerin}, \citenamefont {Houvenaghel}, \citenamefont {Coish}, \citenamefont {Carusotto},\ and\ \citenamefont {St-Jean}}]{pellerin_wavefunction_2024}%
  \BibitemOpen
  \bibfield  {author} {\bibinfo {author} {\bibfnamefont {F.}~\bibnamefont {Pellerin}}, \bibinfo {author} {\bibfnamefont {R.}~\bibnamefont {Houvenaghel}}, \bibinfo {author} {\bibfnamefont {W.~A.}\ \bibnamefont {Coish}}, \bibinfo {author} {\bibfnamefont {I.}~\bibnamefont {Carusotto}},\ and\ \bibinfo {author} {\bibfnamefont {P.}~\bibnamefont {St-Jean}},\ }\bibfield  {title} {\bibinfo {title} {Wave-function tomography of topological dimer chains with long-range couplings},\ }\href {https://doi.org/10.1103/PhysRevLett.132.183802} {\bibfield  {journal} {\bibinfo  {journal} {Phys. Rev. Lett.}\ }\textbf {\bibinfo {volume} {132}},\ \bibinfo {pages} {183802} (\bibinfo {year} {2024})}\BibitemShut {NoStop}%
\bibitem [{\citenamefont {Dutt}\ \emph {et~al.}(2019)\citenamefont {Dutt}, \citenamefont {Minkov}, \citenamefont {Lin}, \citenamefont {Yuan}, \citenamefont {Miller},\ and\ \citenamefont {Fan}}]{dutt_experimental_2019-1}%
  \BibitemOpen
  \bibfield  {author} {\bibinfo {author} {\bibfnamefont {A.}~\bibnamefont {Dutt}}, \bibinfo {author} {\bibfnamefont {M.}~\bibnamefont {Minkov}}, \bibinfo {author} {\bibfnamefont {Q.}~\bibnamefont {Lin}}, \bibinfo {author} {\bibfnamefont {L.}~\bibnamefont {Yuan}}, \bibinfo {author} {\bibfnamefont {D.~A.~B.}\ \bibnamefont {Miller}},\ and\ \bibinfo {author} {\bibfnamefont {S.}~\bibnamefont {Fan}},\ }\bibfield  {title} {\bibinfo {title} {Experimental band structure spectroscopy along a synthetic dimension},\ }\href {https://doi.org/10.1038/s41467-019-11117-9} {\bibfield  {journal} {\bibinfo  {journal} {Nature Communications}\ }\textbf {\bibinfo {volume} {10}},\ \bibinfo {pages} {3122} (\bibinfo {year} {2019})}\BibitemShut {NoStop}%
\bibitem [{\citenamefont {Longhi}(2018)}]{longhi_probing_2018}%
  \BibitemOpen
  \bibfield  {author} {\bibinfo {author} {\bibfnamefont {S.}~\bibnamefont {Longhi}},\ }\bibfield  {title} {\bibinfo {title} {Probing one-dimensional topological phases in waveguide lattices with broken chiral symmetry},\ }\href {https://doi.org/10.1364/OL.43.004639} {\bibfield  {journal} {\bibinfo  {journal} {Opt. Lett.}\ }\textbf {\bibinfo {volume} {43}},\ \bibinfo {pages} {4639} (\bibinfo {year} {2018})}\BibitemShut {NoStop}%
\bibitem [{\citenamefont {Jiao}\ \emph {et~al.}(2021)\citenamefont {Jiao}, \citenamefont {Longhi}, \citenamefont {Wang}, \citenamefont {Gao}, \citenamefont {Zhou}, \citenamefont {Wang}, \citenamefont {Fu}, \citenamefont {Wang}, \citenamefont {Ren}, \citenamefont {Qiao},\ and\ \citenamefont {Jin}}]{jiao_experimentally_2021}%
  \BibitemOpen
  \bibfield  {author} {\bibinfo {author} {\bibfnamefont {Z.-Q.}\ \bibnamefont {Jiao}}, \bibinfo {author} {\bibfnamefont {S.}~\bibnamefont {Longhi}}, \bibinfo {author} {\bibfnamefont {X.-W.}\ \bibnamefont {Wang}}, \bibinfo {author} {\bibfnamefont {J.}~\bibnamefont {Gao}}, \bibinfo {author} {\bibfnamefont {W.-H.}\ \bibnamefont {Zhou}}, \bibinfo {author} {\bibfnamefont {Y.}~\bibnamefont {Wang}}, \bibinfo {author} {\bibfnamefont {Y.-X.}\ \bibnamefont {Fu}}, \bibinfo {author} {\bibfnamefont {L.}~\bibnamefont {Wang}}, \bibinfo {author} {\bibfnamefont {R.-J.}\ \bibnamefont {Ren}}, \bibinfo {author} {\bibfnamefont {L.-F.}\ \bibnamefont {Qiao}},\ and\ \bibinfo {author} {\bibfnamefont {X.-M.}\ \bibnamefont {Jin}},\ }\bibfield  {title} {\bibinfo {title} {Experimentally detecting quantized zak phases without chiral symmetry in photonic lattices},\ }\href {https://doi.org/10.1103/PhysRevLett.127.147401} {\bibfield  {journal} {\bibinfo  {journal} {Phys. Rev. Lett.}\ }\textbf {\bibinfo {volume} {127}},\ \bibinfo {pages}
  {147401} (\bibinfo {year} {2021})}\BibitemShut {NoStop}%
\bibitem [{\citenamefont {Hügel}\ and\ \citenamefont {Paredes}(2014)}]{hugel_chiral_2014}%
  \BibitemOpen
  \bibfield  {author} {\bibinfo {author} {\bibfnamefont {D.}~\bibnamefont {Hügel}}\ and\ \bibinfo {author} {\bibfnamefont {B.}~\bibnamefont {Paredes}},\ }\bibfield  {title} {\bibinfo {title} {Chiral ladders and the edges of quantum {Hall} insulators},\ }\href {https://doi.org/10.1103/PhysRevA.89.023619} {\bibfield  {journal} {\bibinfo  {journal} {Physical Review A}\ }\textbf {\bibinfo {volume} {89}},\ \bibinfo {pages} {023619} (\bibinfo {year} {2014})}\BibitemShut {NoStop}%
\bibitem [{\citenamefont {Vidal}\ \emph {et~al.}(1998)\citenamefont {Vidal}, \citenamefont {Mosseri},\ and\ \citenamefont {Dou\ifmmode~\mbox{\c{c}}\else \c{c}\fi{}ot}}]{vidal_aharanov_1998}%
  \BibitemOpen
  \bibfield  {author} {\bibinfo {author} {\bibfnamefont {J.}~\bibnamefont {Vidal}}, \bibinfo {author} {\bibfnamefont {R.}~\bibnamefont {Mosseri}},\ and\ \bibinfo {author} {\bibfnamefont {B.}~\bibnamefont {Dou\ifmmode~\mbox{\c{c}}\else \c{c}\fi{}ot}},\ }\bibfield  {title} {\bibinfo {title} {Aharonov-bohm cages in two-dimensional structures},\ }\href {https://doi.org/10.1103/PhysRevLett.81.5888} {\bibfield  {journal} {\bibinfo  {journal} {Phys. Rev. Lett.}\ }\textbf {\bibinfo {volume} {81}},\ \bibinfo {pages} {5888} (\bibinfo {year} {1998})}\BibitemShut {NoStop}%
\bibitem [{\citenamefont {Andersson}\ and\ \citenamefont {\"Ostlund}(2003)}]{anderson_staggered_2003}%
  \BibitemOpen
  \bibfield  {author} {\bibinfo {author} {\bibfnamefont {M.}~\bibnamefont {Andersson}}\ and\ \bibinfo {author} {\bibfnamefont {S.}~\bibnamefont {\"Ostlund}},\ }\bibfield  {title} {\bibinfo {title} {Staggered flux and stripes in doped antiferromagnets},\ }\href {https://doi.org/10.1103/PhysRevB.67.014403} {\bibfield  {journal} {\bibinfo  {journal} {Phys. Rev. B}\ }\textbf {\bibinfo {volume} {67}},\ \bibinfo {pages} {014403} (\bibinfo {year} {2003})}\BibitemShut {NoStop}%
\bibitem [{\citenamefont {Aidelsburger}\ \emph {et~al.}(2013{\natexlab{b}})\citenamefont {Aidelsburger}, \citenamefont {Atala}, \citenamefont {Nascimbène}, \citenamefont {Trotzky}, \citenamefont {Chen},\ and\ \citenamefont {Bloch}}]{aidelsburger_experimental_2013}%
  \BibitemOpen
  \bibfield  {author} {\bibinfo {author} {\bibfnamefont {M.}~\bibnamefont {Aidelsburger}}, \bibinfo {author} {\bibfnamefont {M.}~\bibnamefont {Atala}}, \bibinfo {author} {\bibfnamefont {S.}~\bibnamefont {Nascimbène}}, \bibinfo {author} {\bibfnamefont {S.}~\bibnamefont {Trotzky}}, \bibinfo {author} {\bibfnamefont {Y.-A.}\ \bibnamefont {Chen}},\ and\ \bibinfo {author} {\bibfnamefont {I.}~\bibnamefont {Bloch}},\ }\bibfield  {title} {\bibinfo {title} {Experimental realization of strong effective magnetic fields in optical superlattice potentials},\ }\href {https://doi.org/10.1007/s00340-013-5418-1} {\bibfield  {journal} {\bibinfo  {journal} {Appl. Phys. B}\ }\textbf {\bibinfo {volume} {113}},\ \bibinfo {pages} {1} (\bibinfo {year} {2013}{\natexlab{b}})}\BibitemShut {NoStop}%
\bibitem [{\citenamefont {Struck}\ \emph {et~al.}(2013)\citenamefont {Struck}, \citenamefont {Weinberg}, \citenamefont {{\"O}lschl{\"a}ger}, \citenamefont {Windpassinger}, \citenamefont {Simonet}, \citenamefont {Sengstock}, \citenamefont {H{\"o}ppner}, \citenamefont {Hauke}, \citenamefont {Eckardt}, \citenamefont {Lewenstein},\ and\ \citenamefont {Mathey}}]{struck_engineering_2013}%
  \BibitemOpen
  \bibfield  {author} {\bibinfo {author} {\bibfnamefont {J.}~\bibnamefont {Struck}}, \bibinfo {author} {\bibfnamefont {M.}~\bibnamefont {Weinberg}}, \bibinfo {author} {\bibfnamefont {C.}~\bibnamefont {{\"O}lschl{\"a}ger}}, \bibinfo {author} {\bibfnamefont {P.}~\bibnamefont {Windpassinger}}, \bibinfo {author} {\bibfnamefont {J.}~\bibnamefont {Simonet}}, \bibinfo {author} {\bibfnamefont {K.}~\bibnamefont {Sengstock}}, \bibinfo {author} {\bibfnamefont {R.}~\bibnamefont {H{\"o}ppner}}, \bibinfo {author} {\bibfnamefont {P.}~\bibnamefont {Hauke}}, \bibinfo {author} {\bibfnamefont {A.}~\bibnamefont {Eckardt}}, \bibinfo {author} {\bibfnamefont {M.}~\bibnamefont {Lewenstein}},\ and\ \bibinfo {author} {\bibfnamefont {L.}~\bibnamefont {Mathey}},\ }\bibfield  {title} {\bibinfo {title} {Engineering ising-xy spin-models in a triangular lattice using tunable artificial gauge fields},\ }\href {https://doi.org/10.1038/nphys2750} {\bibfield  {journal} {\bibinfo  {journal} {Nature Physics}\ }\textbf {\bibinfo {volume} {9}},\
  \bibinfo {pages} {738} (\bibinfo {year} {2013})}\BibitemShut {NoStop}%
\bibitem [{\citenamefont {Jotzu}\ \emph {et~al.}(2014)\citenamefont {Jotzu}, \citenamefont {Messer}, \citenamefont {Desbuquois}, \citenamefont {Lebrat}, \citenamefont {Uehlinger}, \citenamefont {Greif},\ and\ \citenamefont {Esslinger}}]{jotzu_experimental_2014}%
  \BibitemOpen
  \bibfield  {author} {\bibinfo {author} {\bibfnamefont {G.}~\bibnamefont {Jotzu}}, \bibinfo {author} {\bibfnamefont {M.}~\bibnamefont {Messer}}, \bibinfo {author} {\bibfnamefont {R.}~\bibnamefont {Desbuquois}}, \bibinfo {author} {\bibfnamefont {M.}~\bibnamefont {Lebrat}}, \bibinfo {author} {\bibfnamefont {T.}~\bibnamefont {Uehlinger}}, \bibinfo {author} {\bibfnamefont {D.}~\bibnamefont {Greif}},\ and\ \bibinfo {author} {\bibfnamefont {T.}~\bibnamefont {Esslinger}},\ }\bibfield  {title} {\bibinfo {title} {Experimental realization of the topological haldane model with ultracold fermions},\ }\href {https://doi.org/10.1038/nature13915} {\bibfield  {journal} {\bibinfo  {journal} {Nature}\ }\textbf {\bibinfo {volume} {515}},\ \bibinfo {pages} {237} (\bibinfo {year} {2014})}\BibitemShut {NoStop}%
\bibitem [{\citenamefont {Le}\ \emph {et~al.}(2024)\citenamefont {Le}, \citenamefont {Zhang}, \citenamefont {Cui}, \citenamefont {Wu},\ and\ \citenamefont {Chiu}}]{le_double_2024}%
  \BibitemOpen
  \bibfield  {author} {\bibinfo {author} {\bibfnamefont {C.}~\bibnamefont {Le}}, \bibinfo {author} {\bibfnamefont {Q.}~\bibnamefont {Zhang}}, \bibinfo {author} {\bibfnamefont {F.}~\bibnamefont {Cui}}, \bibinfo {author} {\bibfnamefont {X.}~\bibnamefont {Wu}},\ and\ \bibinfo {author} {\bibfnamefont {C.-K.}\ \bibnamefont {Chiu}},\ }\bibfield  {title} {\bibinfo {title} {Double and quadruple flat bands tuned by alternative magnetic fluxes in twisted bilayer graphene},\ }\href {https://doi.org/10.1103/PhysRevLett.132.246401} {\bibfield  {journal} {\bibinfo  {journal} {Phys. Rev. Lett.}\ }\textbf {\bibinfo {volume} {132}},\ \bibinfo {pages} {246401} (\bibinfo {year} {2024})}\BibitemShut {NoStop}%
\bibitem [{\citenamefont {Senanian}\ \emph {et~al.}(2023)\citenamefont {Senanian}, \citenamefont {Wright}, \citenamefont {Wade}, \citenamefont {Doyle},\ and\ \citenamefont {McMahon}}]{senanian_programmable_2023}%
  \BibitemOpen
  \bibfield  {author} {\bibinfo {author} {\bibfnamefont {A.}~\bibnamefont {Senanian}}, \bibinfo {author} {\bibfnamefont {L.~G.}\ \bibnamefont {Wright}}, \bibinfo {author} {\bibfnamefont {P.~F.}\ \bibnamefont {Wade}}, \bibinfo {author} {\bibfnamefont {H.~K.}\ \bibnamefont {Doyle}},\ and\ \bibinfo {author} {\bibfnamefont {P.~L.}\ \bibnamefont {McMahon}},\ }\bibfield  {title} {\bibinfo {title} {Programmable large-scale simulation of bosonic transport in optical synthetic frequency lattices},\ }\href@noop {} {\bibfield  {journal} {\bibinfo  {journal} {Nat. Phys.}\ }\textbf {\bibinfo {volume} {19}},\ \bibinfo {pages} {1333} (\bibinfo {year} {2023})}\BibitemShut {NoStop}%
\bibitem [{\citenamefont {Cheng}\ \emph {et~al.}(2023)\citenamefont {Cheng}, \citenamefont {Lustig}, \citenamefont {Wang},\ and\ \citenamefont {Fan}}]{cheng_multidimensional_2023}%
  \BibitemOpen
  \bibfield  {author} {\bibinfo {author} {\bibfnamefont {D.}~\bibnamefont {Cheng}}, \bibinfo {author} {\bibfnamefont {E.}~\bibnamefont {Lustig}}, \bibinfo {author} {\bibfnamefont {K.}~\bibnamefont {Wang}},\ and\ \bibinfo {author} {\bibfnamefont {S.}~\bibnamefont {Fan}},\ }\bibfield  {title} {\bibinfo {title} {Multi-dimensional band structure spectroscopy in the synthetic frequency dimension},\ }\href {https://doi.org/10.1038/s41377-023-01196-1} {\bibfield  {journal} {\bibinfo  {journal} {Light: Science {\&} Applications}\ }\textbf {\bibinfo {volume} {12}},\ \bibinfo {pages} {158} (\bibinfo {year} {2023})}\BibitemShut {NoStop}%
\bibitem [{\citenamefont {Dutt}\ \emph {et~al.}(2022)\citenamefont {Dutt}, \citenamefont {Yuan}, \citenamefont {Yang}, \citenamefont {Wang}, \citenamefont {Buddhiraju}, \citenamefont {Vučković},\ and\ \citenamefont {Fan}}]{dutt_creating_2022}%
  \BibitemOpen
  \bibfield  {author} {\bibinfo {author} {\bibfnamefont {A.}~\bibnamefont {Dutt}}, \bibinfo {author} {\bibfnamefont {L.}~\bibnamefont {Yuan}}, \bibinfo {author} {\bibfnamefont {K.~Y.}\ \bibnamefont {Yang}}, \bibinfo {author} {\bibfnamefont {K.}~\bibnamefont {Wang}}, \bibinfo {author} {\bibfnamefont {S.}~\bibnamefont {Buddhiraju}}, \bibinfo {author} {\bibfnamefont {J.}~\bibnamefont {Vučković}},\ and\ \bibinfo {author} {\bibfnamefont {S.}~\bibnamefont {Fan}},\ }\bibfield  {title} {\bibinfo {title} {Creating boundaries along a synthetic frequency dimension},\ }\href {https://doi.org/10.1038/s41467-022-31140-7} {\bibfield  {journal} {\bibinfo  {journal} {Nat Commun}\ }\textbf {\bibinfo {volume} {13}},\ \bibinfo {pages} {3377} (\bibinfo {year} {2022})}\BibitemShut {NoStop}%
\bibitem [{\citenamefont {Chénier}\ \emph {et~al.}(2024)\citenamefont {Chénier}, \citenamefont {d'Aligny}, \citenamefont {Pellerin}, \citenamefont {Édouard Blanchard}, \citenamefont {Ozawa}, \citenamefont {Carusotto},\ and\ \citenamefont {St-Jean}}]{chenier_quantized_2024}%
  \BibitemOpen
  \bibfield  {author} {\bibinfo {author} {\bibfnamefont {A.}~\bibnamefont {Chénier}}, \bibinfo {author} {\bibfnamefont {B.}~\bibnamefont {d'Aligny}}, \bibinfo {author} {\bibfnamefont {F.}~\bibnamefont {Pellerin}}, \bibinfo {author} {\bibfnamefont {P.}~\bibnamefont {Édouard Blanchard}}, \bibinfo {author} {\bibfnamefont {T.}~\bibnamefont {Ozawa}}, \bibinfo {author} {\bibfnamefont {I.}~\bibnamefont {Carusotto}},\ and\ \bibinfo {author} {\bibfnamefont {P.}~\bibnamefont {St-Jean}},\ }\href {https://arxiv.org/abs/2412.04347} {\bibinfo {title} {Quantized hall drift in a frequency-encoded photonic chern insulator}} (\bibinfo {year} {2024}),\ \Eprint {https://arxiv.org/abs/2412.04347} {arXiv:2412.04347 [physics.optics]} \BibitemShut {NoStop}%
\bibitem [{\citenamefont {Dinh}\ \emph {et~al.}(2024)\citenamefont {Dinh}, \citenamefont {Bal{\v{c}}ytis}, \citenamefont {Ozawa}, \citenamefont {Ota}, \citenamefont {Ren}, \citenamefont {Baba}, \citenamefont {Iwamoto}, \citenamefont {Mitchell},\ and\ \citenamefont {Nguyen}}]{dinh_reconfigurable_2024}%
  \BibitemOpen
  \bibfield  {author} {\bibinfo {author} {\bibfnamefont {H.~X.}\ \bibnamefont {Dinh}}, \bibinfo {author} {\bibfnamefont {A.}~\bibnamefont {Bal{\v{c}}ytis}}, \bibinfo {author} {\bibfnamefont {T.}~\bibnamefont {Ozawa}}, \bibinfo {author} {\bibfnamefont {Y.}~\bibnamefont {Ota}}, \bibinfo {author} {\bibfnamefont {G.}~\bibnamefont {Ren}}, \bibinfo {author} {\bibfnamefont {T.}~\bibnamefont {Baba}}, \bibinfo {author} {\bibfnamefont {S.}~\bibnamefont {Iwamoto}}, \bibinfo {author} {\bibfnamefont {A.}~\bibnamefont {Mitchell}},\ and\ \bibinfo {author} {\bibfnamefont {T.~G.}\ \bibnamefont {Nguyen}},\ }\bibfield  {title} {\bibinfo {title} {Reconfigurable synthetic dimension frequency lattices in an integrated lithium niobate ring cavity},\ }\href {https://doi.org/10.1038/s42005-024-01676-9} {\bibfield  {journal} {\bibinfo  {journal} {Communications Physics}\ }\textbf {\bibinfo {volume} {7}},\ \bibinfo {pages} {185} (\bibinfo {year} {2024})}\BibitemShut {NoStop}%
\bibitem [{\citenamefont {Balčytis}\ \emph {et~al.}(2022)\citenamefont {Balčytis}, \citenamefont {Ozawa}, \citenamefont {Ota}, \citenamefont {Iwamoto}, \citenamefont {Maeda},\ and\ \citenamefont {Baba}}]{balcytis_synthetic_2022}%
  \BibitemOpen
  \bibfield  {author} {\bibinfo {author} {\bibfnamefont {A.}~\bibnamefont {Balčytis}}, \bibinfo {author} {\bibfnamefont {T.}~\bibnamefont {Ozawa}}, \bibinfo {author} {\bibfnamefont {Y.}~\bibnamefont {Ota}}, \bibinfo {author} {\bibfnamefont {S.}~\bibnamefont {Iwamoto}}, \bibinfo {author} {\bibfnamefont {J.}~\bibnamefont {Maeda}},\ and\ \bibinfo {author} {\bibfnamefont {T.}~\bibnamefont {Baba}},\ }\bibfield  {title} {\bibinfo {title} {Synthetic dimension band structures on a {Si} {CMOS} photonic platform},\ }\href {https://doi.org/10.1126/sciadv.abk0468} {\bibfield  {journal} {\bibinfo  {journal} {Science Advances}\ }\textbf {\bibinfo {volume} {8}},\ \bibinfo {pages} {eabk0468} (\bibinfo {year} {2022})}\BibitemShut {NoStop}%
\end{thebibliography}%

\end{document}